\documentclass[11pt]{article}
\pdfoutput = 1

\usepackage[utf8]{inputenc}
\usepackage{color,graphicx}
\usepackage{epsfig}
\usepackage{amsmath}
\usepackage{verbatim}
\usepackage{amssymb}
\usepackage{mathabx}
\usepackage[mathcal]{eucal}
\usepackage{stmaryrd}
\usepackage{empheq}
\usepackage{esint}
\usepackage{physics}
\usepackage{amsfonts}
\usepackage{array}
\usepackage{setspace}
\usepackage{braket}
\usepackage{float}
\usepackage{url}
\usepackage{mathtools}
\usepackage{tensor}
\usepackage{bbold}
\usepackage{slashed}
\usepackage{tikz}
\usepackage{graphicx}
\usepackage{epstopdf}
\usepackage{subcaption}
\usepackage[labelfont=bf,font={sf}]{caption}
\usepackage{pgfplots}
\usepackage{mathrsfs}
\usepackage{fancybox}
\usepackage{eurosym}
\usepackage{tcolorbox}
\usepackage{tensor}
\usepackage[normalem]{ulem}
\allowdisplaybreaks
\usepackage{changepage}

\usepackage[margin = 2.2cm]{geometry}
\usepackage[ragged]{footmisc}
\usepackage[bookmarks=true,bookmarksnumbered=false,
hyperindex=true,bookmarksopen=true,hyperfigures=true,
colorlinks=true,linkcolor=webblue,citecolor=webgreen,urlcolor=webblue,breaklinks]{hyperref}
\definecolor{webgreen}{rgb}{0, 0.5, 0}
\definecolor{webblue}{rgb}{0, 0, 0.5}
\definecolor{webred}{rgb}{0.5, 0, 0}
\definecolor{darkgreen}{rgb}{0,0.5,0}

\newcommand{\bmu}{\overline{\mu}}

\newcommand{\average}[1]{\left\langle #1 \right\rangle}

\newcommand{\mjt}{m_\text{JT}}

\def\ben{\begin{equation}}
\def\een{\end{equation}}

   \let\d=\delta 
   
     \let\r=v

\def\be{\begin{equation}}
\def\ee{\end{equation}}
\def\ba{\begin{array}}
\def\ea{\end{array}}

\def\dalemb#1#2{{\vbox{\hrule height .#2pt
       \hbox{\vrule width.#2pt height#1pt \kern#1pt
               \vrule width.#2pt}
       \hrule height.#2pt}}}

\newcommand{\bea}{\begin{eqnarray}}
\newcommand{\eea}{\end{eqnarray}}

\renewcommand{\d}{\mathrm{d}}
\renewcommand{\i}{\mathrm{i}}

\numberwithin{equation}{section}

\begin{document}

\thispagestyle{empty}
    ~\vspace{5mm}
\begin{adjustwidth}{-1cm}{-1cm}
\begin{center}
     {\LARGE \bf 
   Wormholes, branes and finite matrices in sine dilaton gravity}
    
   \vspace{0.4in}

     {\bf Andreas Blommaert$^{1}$, Adam Levine$^2$, Thomas G. Mertens$^3$, Jacopo Papalini$^3$, Klaas Parmentier$^4$}
     \end{center}
    \end{adjustwidth}
\begin{center}
    \vspace{0.4in}
    {$^1$School of Natural Sciences, Institute for Advanced Study, Princeton, NJ 08540, USA\\
    $^2$Center for Theoretical Physics, Massachusetts Institute of Technology,\\ Cambridge, MA 02139, USA\\
    $^3$Department of Physics and Astronomy, Ghent University,\\
    Krijgslaan, 281-S9, 9000 Gent, Belgium\\
    $^4$Department of Physics, Columbia University, New York, NY 10027, USA}
    \vspace{0.1in}
    
    {\tt blommaert@ias.edu, thomas.mertens@ugent.be, 	jacopo.papalini@ugent.be, arlevine@mit.edu, k.parmentier@columbia.edu}
\end{center}

\vspace{0.4in}

\begin{abstract}
\noindent We compute the double trumpet in sine dilaton gravity via WdW quantization. The wormhole size is discretized. The wormhole amplitude matches the spectral correlation of a finite-cut matrix integral, where matrices have large but finite dimensions. This strongly suggests an identification of the sine dilaton gravity theory with the q-deformed JT gravity matrix integral. At the very least, it captures all universal content of that matrix model. The disk decomposes into the physical (gauge invariant) solutions of the WdW equation, which are trumpets with discrete sizes. This decomposition modifies the usual no-boundary wavefunction to a normalizable one in sine dilaton gravity.

We furthermore present an exact quantization of sine dilaton gravity with open and closed end of the world branes. These EOW branes correspond with FZZT branes for the two Liouville theories that make up sine dilaton gravity. The WdW equation implies redundancies in this space of branes, leaving a one parameter family of gauge invariant branes. One gauge choice corresponds with branes discussed by Okuyama in the context of DSSYK. Legendre transforming the EOW brane amplitude reproduces the trumpet. One could read our work as fleshing out the Hilbert space of closed universes in sine dilaton gravity.
\end{abstract}

\pagebreak
\setcounter{page}{1}
\setcounter{tocdepth}{2}
\tableofcontents

\section{Introduction}\label{sect:intro}
In previous work, we introduced and studied a simple model of 2d dilaton gravity \cite{Grumiller:2002nm,Nojiri:2000ja}, which we called sine dilaton gravity \cite{Blommaert:2023opb,Blommaert:2023wad,Blommaert:2024ydx,Blommaert:2024whf,Bossi:2024ffa}
\begin{equation}\label{0.1actionintro}
    I=\frac12\int \mathrm{d} x \sqrt{g}\,(\Phi R+2\sin(\Phi))+\int \mathrm{d}\tau \sqrt{h}\,\Phi K-\mathrm{i}\int\d\tau \sqrt{h} \,e^{-\mathrm{i} \Phi/2}\,.
\end{equation}
We proved using canonical quantization that this model on the disk topology is dual to double-scaled SYK \cite{Cotler:2016fpe,Berkooz:2018jqr,Berkooz:2018qkz}. A salient feature of sine dilaton gravity is that this is the simplest theory with a periodic dilaton potential, invariant under $\Phi\to \Phi+2\pi$. As shown in \cite{Blommaert:2024whf}, this leads universally (for any periodic potential) to a Hamiltonian in which the momentum conjugate to the length of two-sided Cauchy slices is periodic. By consequence, the length of the ERB in periodic dilaton gravity is discrete. An important consequence is that the spectrum is in some sense UV complete - by which we mean that there is a maximal energy:
\begin{equation}
\label{1.2pic}
    \begin{tikzpicture}[baseline={([yshift=-.5ex]current bounding box.center)}, scale=0.7]
    \pgftext{\includegraphics[scale=1]{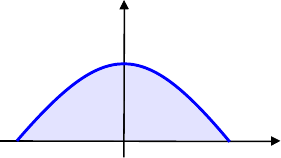}} at (0,0);
    \draw (-2.7,0) node {\color{blue}$\rho_\text{disk}(E)$};
    \draw (-2.2,-1.5) node {$E_\text{min}$};
    \draw (1.5,-1.5) node {$E_\text{max}$};
    \draw (2.2,-0.5) node {$E$};
  \end{tikzpicture}
\end{equation}
This UV completeness is the stand-out new feature of DSSYK as compared to the low energy Schwarzian limit \cite{Maldacena:2016hyu}, and thus maps to periodicity and discrete lengths in the bulk sine dilaton gravity description.

Arguably the main motivation for studying DSSYK, is to learn lessons about quantum cosmology. The hope is that the full SYK model (at large, but finite $N$) has an interpretation in some regime as a microscopic hologram of a cosmological universe. In a 2d cosmological context, as we briefly discuss, the UV completeness maps to a normalizable wavefunction of the universe and replaces the no-boundary theory of initial conditions \cite{hartle1983wave}. The details on the 2d quantum cosmology interpretation of sine dilaton gravity (and of DSSYK) are reported in \cite{Blommaert:2025rgw}.\footnote{There is also significant progress on pinpointing the 3d dS interpretation of sine dilaton gravity, and of DSSYK \cite{Gaiotto:2024kze,Verlinde:2024znh}. Related interesting work is \cite{Collier:2025lux}, although their quantization methodology differs, and their model is therefore not directly related with DSSYK. See section 7.2 of \cite{Blommaert:2024whf} for an explanation of these differences in quantization. See also \cite{Susskind:2022bia,Aguilar-Gutierrez:2024oea}.} Since microscopic holograms of the universe do not grow on trees, we believe that any lessons which can be learned from them should be taken as a strong hint of how to define reasonable theories of quantum gravity. For this reason, we will take sine dilaton gravity, its redundancies in canonical quantization, and the associated discreteness of spacetime seriously, and see where it leads.

\subsection{Summary and structure}

In this paper, we report progress in that direction. Our main result is the calculation of the wormhole amplitude (i.e. the double trumpet \cite{Saad:2019lba}) in sine dilaton gravity via WdW canonical quantization. The relevant Cauchy slices are closed universes. Like in the disk quantization, the conjugate of the size of the universe (which is the horizon value of the dilaton \eqref{3.21} $\Phi_h$) is periodic, and so the universe size is discretized\footnote{We introduce the quantization parameter $\hbar$ in the action which in terms of DSSYK defining variables is $\hbar=2\abs{\log q}$ \cite{Blommaert:2024ydx}.}
\begin{equation}
    \oint_\text{geo}\d s\,e^{-\i \Phi/2}=\hbar b\,,\quad b\in \mathbb{N}_0\,.
\end{equation}
The resulting amplitude encodes the exact spectral correlation expected of a finite-cut matrix integral with leading order DSSYK spectral density
\begin{equation}
    Z_\text{wormhole}(\beta_1,\beta_2)=\sum_{b=1}^\infty b\,I_b(\beta_1/\hbar) I_b(\beta_2/\hbar)=
    \begin{tikzpicture}[baseline={([yshift=-.5ex]current bounding box.center)}, scale=0.7]
    \pgftext{\includegraphics[scale=1]{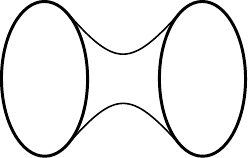}} at (0,0);
    \draw (-2.5,-1) node {$\beta_1$};
    \draw (2.5,-1) node {$\beta_2$};
  \end{tikzpicture}
  \label{1.4zworm}
\end{equation}
This is strong evidence that there ought to be an all-genus definition of sine dilaton gravity that matches with a finite-cut matrix model which was called q-deformed JT gravity in \cite{Jafferis:2022wez}. The WdW quantization of sine dilaton gravity and the associated computation of the wormhole is presented in \textbf{section \ref{sect:trumpets}}. One crucial step is to compute the KG inner product of finite cutoff trumpet wavefunctions, resulting in
\begin{equation}
    \braket{b_1|b_2}=\frac{1}{b_1}\delta_{b_1 b_2}\,.
\end{equation}
This inner product translates to the $b$ measure in \eqref{1.4zworm}. We also show (again using WdW quantization) that this wormhole amplitude \eqref{1.4zworm} is universal for models of dilaton gravity with periodic potentials $V(\Phi+2\pi)=V(\Phi)$ \cite{Blommaert:2024whf}. They thus all plausibly admit finite cut matrix integral interpretations.

For higher genus we can no longer rely on canonical quantization, and we would have to find a path integral definition that is consistent with the Hamiltonian symmetries that we gauged in this work and in \cite{Blommaert:2024whf}. We comment on such a definition in the discussion \textbf{section \ref{sect:concl}}, leaving a detailed investigation to future work. We do emphasize that, aside from the disk (and in a cosmological context the sphere), the wormhole is physically the most important contribution in quantum gravity. Indeed, wormholes encode \emph{all} universal content of matrix models. The sine kernel of eigenvalue correlation \cite{Haake:1315494} is derived just with knowledge of the disk and wormhole (via D-brane logic \cite{Saad:2019lba,Blommaert:2019wfy,Altland:2020ccq,Maldacena:2004sn}). This is more universal than matrix models (whose raison d'\^etre was precisely to have simple models which have this universality).\footnote{In turn, the sine kernel predicts certain universal features of higher genus amplitudes in a particular double scaling limit \cite{Okuyama:2020ncd,Okuyama:2018gfr,Saad:2022kfe,Blommaert:2022lbh,Griguolo:2023jyy}). So, somewhat counterintuitively the disk and wormhole ``know'' features of higher genus amplitudes already.}
The wormhole is also an important contribution in 2d quantum cosmology to the density matrix of the universe, perhaps more important than the disks \cite{Fumagalli:2024msi}.

We stress that this wormhole amplitude is not encoded in DSSYK. Indeed, in DSSYK one genuinely takes $N\to \infty$. As the entropy of the gravity system is of order $N$, higher topologies would be suppressed non-perturbatively in $N$. Relatedly, the exact answers in DSSYK \cite{Berkooz:2018jqr,Berkooz:2018qkz} are already reproduced by sine dilaton gravity at disk level \cite{Blommaert:2024ydx}. One could think of our wormhole amplitude as closer to ordinary large $N$ SYK, which indeed has a ramp \cite{Cotler:2016fpe} consistent with our wormhole. 

In the second part of this paper, we will study the space of EOW branes \cite{Kourkoulou:2017zaj} in sine dilaton gravity. For this we consider in \textbf{section \ref{sect:classify}} the theory with action
\begin{align}
    I=\frac{1}{2}\int \d x \sqrt{g}\,(\Phi R+2\sin(\Phi))+\int \d u\sqrt{h}\,\Phi K-\overline{\mu} \int_\text{brane} \d u\sqrt{h}\,e^{\i \Phi/2}-\mu \int_\text{brane} \d u\sqrt{h}\,e^{-\i \Phi/2}\,.
\end{align}
These boundary terms correspond with turning on boundary cosmological constants in the formulation of sine dilaton gravity as two copies of Liouville CFT \cite{Blommaert:2024ydx,Verlinde:2024zrh,Collier:2025pbm}. So, these are FZZT branes \cite{Fateev:2000ik}. We show that for closed EOW branes there are gauge redundancies in this two-parameter space of FZZT branes and that gauge-inequivalent branes are specified by the dilaton at the horizon  
\begin{equation}
    \Phi_h= -\i\text{arcsinh}(\mu) - \i \text{arcsinh}(\overline{\mu})\,,\quad [b,\Phi_h]=\i\,.\label{1.7ginv}
\end{equation}
We study the canonical quantization of sine dilaton gravity in the presence of EOW branes, mimicking the setup of Gao-Jafferis-Kolchmeyer \cite{Gao:2021uro} for JT gravity. A one-parameter family of branes corresponds with the branes discussed by Okuyama \cite{Okuyama:2023byh}. It is often convenient to think about sine dilaton gravity using an effective AdS$_2$ spacetime $\d s^2_\text{AdS}=e^{-\i \Phi} \d s^2$. The FZZT branes then describe trajectories that branes with fixed mass $m_\text{AdS}$ and extrinsic curvature $K_\text{AdS}$ would follow in ordinary JT gravity, with the relation between these two parameters and $(\mu,\overline{\mu})$ given in \eqref{eqn:muGmuMvsKm}.

We believe the study of these EOW branes is interesting in its own right. However, we had ulterior motives. Namely, canonical quantization in the presence of branes allows for an independent derivation of the trumpet amplitude $I_b(\beta/\hbar)$ in sine dilaton gravity, providing a check on our WdW quantization of section 2. We show in \textbf{section \ref{subsect:4.4recover}} that the EOW brane cylinder amplitude equals
\begin{equation}
    Z_\text{EOW}(\Phi_h,\beta)=\sum_{b=0}^\infty \frac{e^{-\i b \Phi_h}}{1-q^{2b}}\,I_b(\beta/\hbar)=
    \begin{tikzpicture}[baseline={([yshift=-.5ex]current bounding box.center)}, scale=0.7]
    \pgftext{\includegraphics[scale=1]{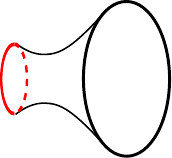}} at (0,0);
    \draw (-2,0) node {\color{red}$\Phi_h$};
    \draw (1.7,-1) node {$\beta$};
  \end{tikzpicture}
\end{equation}
This indeed only depends on the gauge-invariant combination \eqref{1.7ginv}. Fourier transforming to the $b$ basis correctly reproduces the trumpet amplitude.

In \textbf{section \ref{sect:7matrixdictionary}}, we derive the matrix integral dual operator insertions that correspond with inserting trumpet boundaries and FZZT brane boundaries in the gravitational path integral.\footnote{These expressions do not actually assume a matrix integral duality beyond the leading topology, they are valid generically under the assumption that the gravity model is compatible with a certain collection of moments of the spectral density. This does not assume a matrix integral, the equations also work for different statistics such as in the SYK model.} The FZZT branes correspond with the insertion of $\Tr \log(H+\cos(\Phi_h))$. This generalizes $\Tr \log (H-E)$ from usual double scaled matrix integrals (such as minimal strings \cite{Maldacena:2004sn,Collier:2023cyw}) to finite-cut ones.

Finally, in the discussion \textbf{section \ref{sect:concl}}, we suggest how to make sense of sine dilaton gravity with the symmetry $\Phi(x)\to \Phi(x)+2\pi$ on higher genus topologies. The crux is that ``crotch singularity'' operator insertions \cite{Louko:1995jw} (which are in a sense internally provided by the path integral \cite{Blommaert:2023vbz}) can compensate the Euler character deficit that comes from the $\Phi \chi$ term in the action \eqref{0.1actionintro}. This mechanism corresponds with screening charges in the Coulomb gas formalism of generalized minimal strings \cite{Zamolodchikov:2005fy,Dotsenko:1984ad,Belavin:2005yj}, which is closely related with sine dilaton gravity. We also speculate on how the appearance of a normalizable wavefunction of the universe in sine dilaton gravity is related with including quantum observers \cite{Chandrasekaran:2022cip} in cosmology.

\subsection{Background and notation}
Before diving into the WdW quantization, we collect some equations and conventions that are required to follow the main text. For in-depth explanations and background on sine dilaton gravity we refer the readers to  \cite{Blommaert:2023opb,Blommaert:2023wad,Blommaert:2024ydx,Blommaert:2024whf} (with identical conventions). The classical solutions of low Euler character $\abs{\chi}$ are
\begin{equation}
    \d s^2=F(r)\d \tau^2+\frac{1}{F(r)}\d r^2\,,\quad F(r)=-2\cos(r)+2\cos(\theta)\,,\quad \Phi=r\,,\quad \Phi_h=\theta\,.\label{1.9}
\end{equation}
It is often convenient to work with a Weyl rescaled AdS metric $\d s^2_\text{AdS}=e^{-\i \Phi}\d s^2$ such that the solutions become the JT solutions
\begin{equation}
    \d s^2_\text{AdS}=F_\text{AdS}(\rho)\d \tau ^2+\frac{1}{F_\text{AdS}(\rho)}\d \rho^2\,,\quad F_\text{AdS}(\rho)=\rho^2-\sin(\theta)^2\,.\label{solutions}
\end{equation}
The map between the AdS coordinate $\rho=\Phi_\text{AdS}$ and the sine dilaton gravity coordinate $r=\Phi$ is 
\begin{equation}
    e^{-\i \Phi}=\cos(\theta)-\i\Phi_\text{AdS}\,.\label{1.11dilatons}
\end{equation}
The holographic DSSYK boundary conditions are $\sqrt{h}\,e^{\i \Phi/2}=\i$ with $\Phi=\pi/2+\i\infty$, and one could align the complex metric contour with the real $\rho$ contour in the AdS metric \eqref{solutions} (with the hologram at the usual boundary location $\rho=+\infty$). We will use three different length variables for Cauchy slices
\begin{equation}
    \ell=\oint_\text{slice}\d s\,,\quad L_\text{AdS}=\oint_\text{slice}\d s\,e^{-\i \Phi/2}\,,\quad\overline{L}_\text{AdS}=\oint_\text{slice}\d s\,e^{\i \Phi/2}\,.\label{1.26lengths}
\end{equation}
The first is the actual length in sine dilaton gravity. The others are lengths measured in the AdS metric \eqref{solutions} and a second AdS metric (with the conjugate Weyl factor).\footnote{In \cite{Xu:2024gfm, Heller:2024ldz} a quantum mechanical correspondence was established between $L_\text{AdS}$ in sine dilaton gravity and the Krylov spread complexity in DSSYK \cite{Rabinovici:2023yex,Ambrosini:2024sre}.} In section \ref{sect:classify}, we will occasionally use the notation $L$ without additional specification, in which case we always mean the length $L_\text{AdS}$ in the AdS metric. The Hamiltonian which describes two-sided time evolution in sine dilaton gravity is\footnote{Throughout this paper we will mostly use this Hermitian version of the sine dilaton gravity Hamiltonian.}
\begin{equation}
    \mathbf{H}=\frac{1}{2}e^{\i \mathbf{P}}\sqrt{1-e^{-\mathbf{L}}}+\frac{1}{2}\sqrt{1-e^{-\mathbf{L}}}e^{-\i \mathbf{P}}\,,\quad \mathbf{P}\sim \mathbf{P}+2\pi\,,\quad [\mathbf{L},\mathbf{P}]=\i \hbar\,.\label{1.ham}
\end{equation}
This is the chord Hamiltonian \cite{Berkooz:2018jqr,Berkooz:2018qkz,Lin:2022rbf} of DSSYK with parameter mapping
\begin{equation}
    \hbar = 2\abs{\log q}=-2\pi \i b^2_\text{CFT}
\end{equation}
The parameter $b_\text{CFT}$ labels the central charge in the Liouville rewriting of sine dilaton gravity \cite{Blommaert:2024ydx,Verlinde:2024zrh,Collier:2025pbm}. Introducing $\ell=e^\rho$, the two Liouville fields $(\varphi,\overline{\varphi})$ or $(\phi,\chi)$ are
\begin{equation}\label{000}
    \varphi=\rho-\i \Phi/2=\i b_\text{CFT} \chi\,,\quad \overline{\varphi}=\rho+\i \Phi/2=b_\text{CFT}\phi
\end{equation}
Here $b_\text{CFT}$ labels the central charge associated with the Liouville field $\overline{\varphi}$ and the central charge associated with $\varphi$ has the complex conjugate label $\i b_\text{CFT}$.

\section{Wormholes}\label{sect:trumpets}
The purpose of this section is to compute the wormhole amplitude in sine dilaton gravity \cite{Blommaert:2024ydx} with action
\begin{equation}\label{Irescaled}
    I=\frac12\int \mathrm{d} x \sqrt{g}\,(\Phi R+2\sin(\Phi))+\int \mathrm{d}\tau \sqrt{h}\,\Phi K-\mathrm{i}\int\d\tau \sqrt{h} \,e^{-\mathrm{i} \Phi/2}\,.
\end{equation}
The last boundary term is required for holographic normalization. We compute the wormhole amplitude via WdW quantization. In \textbf{section \ref{subsect:wdw}} we discuss the minisuperspace WdW equation. For 2d dilaton gravity, minisuperspace is exact \cite{henneaux1985quantum,Maldacena:2019cbz,Iliesiu:2020zld,Held:2024rmg}. In \textbf{section \ref{subsect:finitecutoff}} we identify the solutions of the WdW equation $H_\text{WdW}=0$ as finite cutoff versions of the trumpet in sine dilaton gravity. In \textbf{section \ref{sect:doubletrumpet}} we compute the norm on these physical states using a KG inner product. Inserting the associated identity operator (or projector) between generic (non-physical) states results in the double trumpet amplitude.

After completing our calculations for this section, a paper appeared \cite{Held:2024rmg} in which several aspects of WdW quantization of 2d dilaton gravity are explained very pedagogically. 

\subsection{WdW equation and closed channel quantization}\label{subsect:wdw}
Variation of the sine dilaton gravity action \eqref{0.1actionintro} (without the last renormalization term) and using the bulk EOM leaves us with the boundary terms
\begin{equation}
    \delta I=\int\d \tau \bigg(n^\mu \partial_\mu \Phi\,\delta \sqrt{h}+\sqrt{h} K\, \delta \Phi\bigg)\,.\label{phasespace}
\end{equation}
One recognizes the two familiar Legendre conjugate pairs of 2d dilaton gravity \cite{Goel:2020yxl}.\footnote{The symplectic form on phase space follows from the on-shell variation of the action
\begin{equation}
    I=\i \int_{t_i}^{t_f}\d t\,\bigg(\sum_{i=1}^n p_i \frac{\d}{\d t} q_i-H(p_i,q_i)\bigg)\,,\quad \delta I=\i \sum_{i=1}^n p_i\,\delta q_i\rvert_{t_f}\,.
\end{equation}} We can either fix $\sqrt{h}$ or the momentum $n^\mu \partial_\mu \Phi$, and in addition we can fix $\Phi$ or the curvature $K$. Working in the ADM formalism and in minisuperspace (zero mode sector), the WdW Hamiltonian is found to be\footnote{See for instance equation (3.1) in \cite{Held:2024rmg} or (2.20) in \cite{Iliesiu:2020zld}. As compared to (2.20) in \cite{Iliesiu:2020zld} we are led to use different operator ordering 
\begin{equation}
    \ell\frac{\d}{\d \ell}\frac{1}{\ell}\to \frac{\d}{\d \ell}\,.
\end{equation}
Our particular choice of operator ordering gets fixed by imposing that our solutions match with the results of 2d Liouville CFT. We believe consequentially that the most reasonable operator ordering to use in JT gravity is also instead \eqref{2.6hwdw}.}
\begin{equation}
    H_\text{WdW}=\hbar^2\frac{\d}{\d \Phi}\frac{\d}{\d \ell}-\sin(\Phi)\,\ell\,,\quad \hbar=2\abs{\log q}\label{2.6hwdw}
\end{equation}
This should be understood as a constraint differential operator action on gauge-invariant wavefunctions of $(\ell,\Phi)$
\begin{equation}
    H_\text{WdW}\,\psi(\ell,\Phi)=0\,.
\end{equation}
In terms of the ADM path integral, the constraint arises because the lapse function is a Lagrange multiplier. The Hamiltonian generates time translations, but those are diffeomorphisms and therefore are to be gauged in gravity. The momentum constraint (associated with gauging spatial diffeomorphisms) trivially vanishes in the minisuperspace approximation (and in the absence of matter particles) \cite{Held:2024rmg,Iliesiu:2020zld}.

We stress that the quantum WdW Hamiltonian \eqref{2.6hwdw} is invariant under periodic shifts of the dilaton zero mode $\Phi\to \Phi+2\pi$. We treat this as a redundancy (we gauge this symmetry) because not doing so leads to divergences in the wormhole amplitude (see section \ref{subsect:ungworm}). In the disk quantization \cite{Blommaert:2024whf}, a similar shift symmetry must indeed also be gauged in order to obtain finite (and thus meaningful) results. As explained in the introduction, for our purposes, this periodicity of the dilaton is the hallmark feature of sine dilaton gravity. So we impose on physical wavefunctions the analogous minisuperspace periodicity
\begin{equation}
    \psi(\ell,\Phi+2\pi)=\psi(\ell,\Phi)\,.\label{2.7}
\end{equation}
We should stress that the action \eqref{Irescaled} only has a symmetry under $\Phi\to \Phi+2\pi$ if the theory is considered on a topology with Euler character $\chi=0$. This is the case for cylinder amplitudes, for which canonical quantization in the closed channel can be used. We speculate in the discussion section \ref{subsect:82generalizedminimal} on how one could potentially enforce $\Phi\sim \Phi+2\pi$ on higher genus surfaces.

The solutions to this constraint represent the physical states of the theory. The pre-Hilbert space is $L^2(\ell)\otimes L^2(\Phi)$.\footnote{For more background see \cite{Held:2024rmg,Witten:2022xxp,Chandrasekaran:2022cip}.} Boundary conditions in 2d gravity should be interpreted as corresponding to states $\ket{\ell,\Phi}$ in this larger space. Gravitational time evolution between slices implements a projection on the subspace $H_\text{WdW}=0$. In the pre-Hilbert space this projector is $\mathbf{1}_\text{phys}=\delta(\mathbf{H}_\text{WdW})$, so that the wormhole amplitude can be decomposed as
\begin{equation}
    Z_\text{wormhole}(\ell_1,\Phi_1,\ell_2,\Phi_2)=\bra{\ell_1,\Phi_1}\mathbf{1}_\text{phys}\ket{\ell_2,\Phi_2}=
    \begin{tikzpicture}[baseline={([yshift=-.5ex]current bounding box.center)}, scale=0.7]
    \pgftext{\includegraphics[scale=1]{sdworm1.pdf}} at (0,0);
    \draw (-2.9,-1) node {$\ell_1,\Phi_1$};
    \draw (2.9,-1) node {$\ell_2,\Phi_2$};
  \end{tikzpicture}
    \label{2.7worm}
\end{equation}
Physical states $\ket{\psi}$ satisfy
\begin{equation}
    \mathbf{H}_\text{WdW}\ket{\psi}=0\,,\quad \psi(\ell,\Phi)=\left\langle \ell,\Phi\rvert \psi\right\rangle\,.
\end{equation}
Our goal in this section is to find a basis of these physical states, their wavefunctions and the associated identity operator $\mathbf{1}_\text{phys}$ so as to compute the wormhole amplitude \eqref{2.7worm}. Equations (5.10) and (6.12) in \cite{Held:2024rmg} admit similar interpretation in terms of the wormhole amplitude \cite{Saad:2019lba} in JT gravity.

We now rewrite the WdW differential equation \eqref{2.6hwdw} in a form that enables a simple exact solution. Introducing the conformal factor of the metric $\ell=e^\rho$, the WdW constraint equation becomes
\begin{equation}
    \hbar^2\frac{\d}{\d \Phi}\frac{\d}{\d\rho}-e^{2\rho}\sin(\Phi)=0\,.
\end{equation}
Further transforming to the zero modes of the two Liouville CFT fields using \eqref{000}, this takes a familiar form related with the stress tensor from Liouville field theory\footnote{See for instance equation (4.2) in \cite{Anninos:2024iwf} for a closely related scenario but with the fixed state of the matter sector replaced by the dynamical second Liouville zero mode variable.}
\begin{equation}
    -\frac{\d^2}{\d \phi^2}+\frac{1}{4\pi^2 b_\text{CFT}^2}e^{2 b_\text{CFT} \phi}-\frac{\d^2}{\d \chi^2}-\frac{1}{4\pi^2 b_\text{CFT}^2}e^{2 \i b_\text{CFT} \chi}=0\,.\label{2.11diffeq}
\end{equation}
We see here already that this is indeed the natural operator ordering associated with a two-dimensional target space (or superspace, in this language) with metric $\d \phi^2+\d\chi^2$.\footnote{We could introduce space and time coordinates $\exp(T)=\exp(\i b_\text{CFT} \chi)/2\abs{\log q}$ and $\exp(X)=\exp(b_\text{CFT} \phi)/2\abs{\log q}$. The result is a time dependent exponential background potential. This is standard in cosmological models (for instance in the Airy model that describes minisuperspace 4d Einstein-Hilbert gravity, see for instance \cite{Caputa:2018asc}).} We solidify this statement by comparing with Liouville CFT disk one point functions at the end of section \ref{subsect:finitecutoff}. The main interesting feature of \eqref{2.11diffeq} is its factorized form so that we can solve it by separation of variables. A final rewriting to the AdS length variables \eqref{1.26lengths} $L_\text{AdS}=e^{\i b_\text{CFT} \chi}$ and $\overline{L}_\text{AdS}=e^{b_\text{CFT}\phi}$ results in the WdW equation
\begin{equation}
    \boxed{\hbar ^2 \overline{L}_\text{AdS}\frac{\d}{\d \overline{L}_\text{AdS}}\overline{L}_\text{AdS}\frac{\d}{\d \overline{L}_\text{AdS}}+\overline{L}_\text{AdS}^2=\hbar ^2 L_\text{AdS}\frac{\d}{\d L_\text{AdS}}L_\text{AdS}\frac{\d}{\d L_\text{AdS}}+L_\text{AdS}^2=\hbar^2 b^2\,}\label{2.24wdwh}
\end{equation}

Before solving this equation, we would like to understand the meaning of the parameter $b$ that labels the solutions of our WdW constraint, and therefore labels states in the physical Hilbert space of sine dilaton gravity. For this, we briefly get ahead of ourselves and use (a fact explained in detail in section \ref{sect:classify}) that the conjugates of the length variables are the respective effective AdS curvatures
\begin{equation}
    K_\text{AdS}=-\i \hbar \frac{\d}{\d L_\text{AdS}}\,,\quad \overline{K}_\text{AdS}=\i \hbar \frac{\d}{\d \overline{L}_\text{AdS}}\,.\label{2.13canstr}
\end{equation}
Furthermore, following \cite{Held:2024rmg}, we introduce $k_\text{AdS}=\i L_\text{AdS} K_\text{AdS}$ such that the WdW constraint equation \eqref{2.24wdwh} becomes simply
\begin{equation}    \overline{k}^2_\text{AdS}+\overline{L}^2_\text{AdS}=k^2_\text{AdS}+L^2_\text{AdS}=\hbar^2 b^2\,.
\end{equation}
The Hamiltonian thus generates rotations in the $(k_\text{AdS},L_\text{AdS})$ plane, idem for the other two coordinates. These are the ordinary radial translations on the AdS$_2$ double trumpet geometry discussed in \cite{Held:2024rmg}. We can classify physical configurations for instance by considering $k_\text{AdS}=0$ slices in the geometry. These are closed AdS geodesics with a wavefunction $\sim \delta(L_\text{AdS}=\hbar b)$. The WdW evolution then generates a more general physical wavefunction $\psi_b(L_\text{AdS},\overline{L}_\text{AdS})$ solving \eqref{2.24wdwh}. Because the latter is simply radial evolution, we generate a finite cutoff trumpet amplitude
\begin{equation}
    \psi_b(\ell,\Phi)=Z_\text{trumpet}(b,\ell,\Phi)=
    \begin{tikzpicture}[baseline={([yshift=-.5ex]current bounding box.center)}, scale=0.7]
    \pgftext{\includegraphics[scale=1]{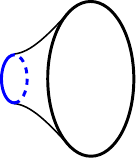}} at (0,0);
    \draw (-1.55,0) node {\color{blue}$b$};
    \draw (1.75,-1) node {$\ell,\Phi$};
  \end{tikzpicture}
\end{equation}
In other words, the eigenvalue $b$ of \eqref{2.24wdwh} is directly related in the full quantum theory with the length of the closed geodesic in the effective AdS geometry 
\begin{equation}
    \boxed{\oint_\text{geo}\d s\,e^{-\i \Phi/2}=\hbar b\,}\label{2.16hbarb}
\end{equation}
This interpretation for the conserved quantity $k_\text{AdS}^2+L_\text{AdS}^2$ was also given in \cite{Held:2024rmg}. The difference with our case is that for sine dilaton gravity, the canonical structure \eqref{2.13canstr} is different from that of ordinary JT gravity (for instance, in JT gravity $k_\text{AdS}$ is conjugate to the dilaton), leading to different finite cutoff wavefunctions.

\subsection{Trumpets}\label{subsect:finitecutoff}
We will now uniquely determine the finite cutoff trumpets in sine dilaton gravity, meaning the solutions to \eqref{2.24wdwh} that satisfy certain boundary conditions for $\ell\to 0$ and $\ell\to \infty$. We claim that a correct solution is
\begin{equation}
    \boxed{Z_\text{trumpet}(b,\ell,\Phi)=\frac{\pi^{1/2}}{2^{3/2}}\,J_b(\ell e^{\i \Phi/2}/\hbar)\,Y_b(\ell e^{-\i \Phi/2}/\hbar)+(\ell\to-\ell)\,,\,b\in \mathbb{N}_0}\label{2.18fctrumpet}
\end{equation}
We now prove this.

The general solution to \eqref{2.24wdwh} is a combination of $J_b(L_\text{AdS}/\hbar)$ and $Y_b(L_\text{AdS}/\hbar)$, likewise for $\overline{L}_\text{AdS}$. We now compare with the wavefunction in two asymptotic limits. Firstly there is the holographic limit. The DSSYK holographic boundary conditions are $\Phi=\pi/2+\i \infty$ and \cite{Blommaert:2024ydx}
\begin{equation}
    \sqrt{h}\,e^{\i \Phi/2}=\i\,.\label{2.19bc}
\end{equation}
This translates to
\begin{equation}
    L_\text{AdS}=+\infty\,,\quad \overline{L}_\text{AdS}=\i \oint \d \tau=\i \beta\,.\label{2.26ass}
\end{equation}
The first equation means that we go to the usual holographic boundary $\rho_\text{AdS}=+\infty$ in the AdS metric. In this limit
\begin{equation}
    Z_\text{trumpet}(b,\ell,\Phi)\to Z_\text{trumpet}(b,\beta)\,,\label{2.27}
\end{equation}
with the asymptotic trumpet given by
\begin{equation}
    \boxed{Z_\text{trumpet}(b,\beta)=I_b(\beta/\hbar)\,}\label{2.28}
\end{equation}
Here we stripped off infinitely fast oscillating factors using a type of holographic normalization. \eqref{2.28} is the correct answer, so this uniquely determines the dependence on $\overline{L}_\text{AdS}$ in \eqref{2.18fctrumpet} to be $J_b(\overline{L}_\text{AdS}/\hbar)$. Why can we claim that \eqref{2.28} is the correct answer? We will recover this same answer from a completely independent calculation in section \ref{subsect:4.4recover}, by first computing the amplitude of a cylinder ending on an EOW brane, and then Fourier transforming the branes - mimicking a calculation by \cite{Gao:2021uro} for JT. An additional (indirect) piece of evidence for \eqref{2.28} is that the disk amplitude of sine dilaton gravity decomposes into the trumpets \eqref{2.28}, as geometrically expected. We comment more on this near the end of this section and in the discussion section \ref{subsect:nonoboundary}. This same expression was found by Okuyama \cite{Okuyama:2023byh}.

Secondly, we can consider the small surface limit $\ell\to 0$, and compute the gravitational path integral on-shell. For disk topologies and $\ell\to 0$
\begin{equation}
    \exp\bigg( \frac{1}{2\hbar}\int \d x \sqrt{g}(\Phi R+2\sin(\Phi))+\frac{1}{\hbar}\int \d u \sqrt{h}K\bigg)\overset{\ell\to 0}{=} e^{2\pi \Phi/\hbar}\,.\label{2.21}
\end{equation}
Indeed, the area goes to zero so the potential term may be ignored, and the dilaton can be approximated as constant. We then recognize the Euler character. Conical defects with opening angle $\gamma$ are obtained by inserting certain bulk operators in the gravitational path integral \cite{Mertens:2019tcm,Maxfield:2020ale}
\begin{equation}
    V_\gamma=\int \d x \sqrt{g}\,e^{-(2\pi-\gamma)\Phi/\hbar}\,.
\end{equation}
The strength of the defect is a local property, and therefore it is the same in the AdS metric \eqref{solutions} as in the actual sine dilaton gravity metric \eqref{1.9}. AdS metrics with imaginary-$\gamma$ conical defects have real geodesics if we analytically continue the metric contour (see for instance appendix A in \cite{Blommaert:2021fob}). So the defect that creates our AdS geodesic boundary \eqref{2.16hbarb} is
\begin{equation}
    V_b=\int \d x \sqrt{g}\,e^{(\i b-2\pi/\hbar)\Phi}\,.\label{2.23}
\end{equation}
Path integrals on tiny surfaces $\ell\to 0$ with such defects inserted are then approximated to leading order in $1/\hbar$ by multiplying the vertex operator with the topological contribution from \eqref{2.21}. This gives the leading order wavefunction $e^{\i b \Phi}$. Taking into account the small argument expansion of the Bessel functions, we determine there must be a piece of the wavefunction proportional to $Y_b(L_\text{AdS}/\hbar)$. Indeed, our finite cutoff trumpets \eqref{2.18fctrumpet} are consistent with the $\ell\to 0$ path integral
\begin{equation}
    Z_\text{trumpet}(b,\ell\to 0,\Phi)=\frac{1}{(2\pi)^{1/2}}\frac{1}{b}e^{\i b \Phi}\,.\label{2.24}
\end{equation}
This should be interpreted as a state-operator correspondence where operators \eqref{2.23} create states with wavefunctions \eqref{2.24}. This state is an eigenstate of the conjugate of $\Phi_h$, with eigenvalues equal to the AdS geodesic lengths:
\begin{equation}
    -\i \hbar \frac{\d}{\d \Phi}=\hbar b\,.
\end{equation}
This leaves the freedom to add to \eqref{2.18fctrumpet} the combination $J_b(L_\text{AdS}/\hbar)J_b(\overline{L}_\text{AdS}/\hbar)$. However, this combination does not contribute to the KG inner product which we compute in section \ref{sect:doubletrumpet}. We interpret this therefore as the addition of a null state and ignore it. The case $b=0$ is special. The combination involving $Y_0(L_\text{AdS}/\hbar)$ diverges logarithmically for $\ell\to 0$ and therefore is not a reasonable wavefunction. This leaves only the contribution $J_0(L_\text{AdS}/\hbar)J_0(\overline{L}_\text{AdS}/\hbar)$. As mentioned previously, via the KG inner product of section \ref{sect:doubletrumpet}, we find that this has zero norm. In conclusion, the state $\ket{b=0}$ is a null state.

Finally we must impose invariance under $\Phi\to \Phi+2\pi$ \eqref{2.7}. This discretizes the conjugate of $\Phi_h$, such that $b\in\mathbb{Z}$. We indeed observe that for $\ell\to 0$ the only gauge-invariant wavefunctions \eqref{2.24} have integer $b$. This is also the reason for including $\ell\to-\ell$ in \eqref{2.18fctrumpet}. The wavefunctions are invariant under $b\to -b$ so we may restrict to positive integers in \eqref{2.18fctrumpet}.

Before proceeding with computing the inner product on these wavefunctions \eqref{2.18fctrumpet}, so as to compute the double trumpet \eqref{2.7worm}, let us make two comments.
\begin{enumerate}
    \item In Liouville gravity (with a real $b_\text{CFT}$) \cite{Mertens:2020hbs, Fan:2021bwt, Blommaert:2023wad}, the solutions of the WdW equation \eqref{2.11diffeq} are
    \begin{equation}
        \psi_\omega(\phi,\chi)=K_{\i \omega}(e^{b_\text{CFT}\phi}/2\pi b_\text{CFT}^2)K_{\i \omega}(e^{\i b_\text{CFT} \chi}/2\pi b_\text{CFT}^2)\,,\quad \omega=\i b\,.\label{2.29}
    \end{equation}
    This is the product of Liouville bulk one point functions \cite{Fateev:2000ik,Mertens:2020hbs} with\footnote{We follow the notation for $P$ of equation (4.7) in \cite{Mertens:2020hbs}.}
    \begin{equation}
        P=\omega \frac{b_\text{CFT}}{2}\,.\label{2.30}
    \end{equation}
    The WdW constraint (or Virasoro constraint or target space on shell relation) forces the momenta $P$ of the two Liouville fields to match. The fact that we find the correct Liouville disk one point functions demonstrates that we used the ``correct'' operator ordering in our WdW Hamiltonian. The Liouville bulk vertex operator with momentum \eqref{2.30} indeed becomes the defect operator $V_b$ \eqref{2.23} using the field redefinition \eqref{000}. We comment further on this Liouville interpretation in the discussion section \ref{subsect:82generalizedminimal}.
    \item The disk amplitude in sine dilaton gravity \cite{Blommaert:2024whf} must be a solution to the WdW equation, meaning that it should be expandable in the basis of solutions \eqref{2.28} (i.e. the trumpet amplitudes). Such a decomposition was indeed found by Okuyama \cite{Okuyama:2024eyf}\footnote{An analogous semiclassical version of this formula was written down in section 4.2 of \cite{Blommaert:2024ydx}. The sum over $b$ may also be limited to run from $0$ to $\infty$, the trumpet \eqref{2.28} is invariant under $b\to -b$.}
    \begin{equation}
    \label{eq:okudecomp}
        Z_\text{disk}(\beta)=\sum_{b=-\infty}^{+\infty}\psi_\text{disk}(b)\, Z_\text{trumpet}(b,\beta)\,.
    \end{equation}
    The wavefunction is
    \begin{equation}
        \psi_\text{disk}(b)=\cos(\pi b/2)q^{b^2/4+b/2}=\i q^{(b+1+\i \pi/\abs{\log q})^2/4}-\i q^{(b+1-\i \pi/\abs{\log q})^2/4}\label{2.32}
    \end{equation}
    These Gaussian densities replace the usual no-boundary wavefunction in 2d dilaton gravity, which, in ``ungauged'' sine dilaton \cite{Blommaert:2024whf} and in Liouville gravity, are delta functions centered around these same values\footnote{See for instance appendix B of \cite{Mertens:2020hbs}, where the disk is identified with a $(1,1)$ ZZ-brane disk one point function, interpreted as the subtraction of the $(1,1)$ and $(1,-1)$ degenerate operators \cite{Martinec:2003ka}. This subtraction explains the minus sign.}
    \begin{equation}
        \psi_\text{disk\,ungauged}(b)=\i \delta(b+1+\i \pi/\abs{\log q})-\i \delta(b+1-\i \pi/\abs{\log q})\,.\label{2.33}
    \end{equation}
    These delta functions can be understood to fix $\ket{b=2\pi i/\hbar}$ which is the usual smooth no-boundary state \cite{hartle1983wave} (the absence of a conical defect \eqref{2.23}).\footnote{The relation with one delta function at $b=2\pi \i/\hbar$ appears when we consider a holographic boundary with an additional marked point \cite{Mertens:2020hbs}. This gives the amplitude $\beta Z_\text{disk}(\beta)$, which indeed in the $b$ basis is a single delta function. In intersection theory, this exact non-perturbative identity is called the string equation, see for instance equation (5.18) in \cite{Blommaert:2022lbh}.} In ``UV complete'' models of quantum gravity (like periodic dilaton gravity) we have a different theory of initial conditions \eqref{2.32} which smears the usual un-normalizable no-boundary state \eqref{2.33}. We comment on interpretations of this new state as associated with including observers in cosmology \cite{Chandrasekaran:2022cip,Abdalla:2025gzn,Harlow:2025pvj} in the discussion section \ref{subsect:nonoboundary}. Details of the 2d cosmological interpretation of sine dilaton gravity will be reported elsewhere.
\end{enumerate}

\subsection{Universal periodic wormholes and finite dimensional random matrices}\label{sect:doubletrumpet}
We would now like to compute the double trumpet using canonical quantization
\begin{equation}
    Z_\text{wormhole}(\ell_1,\Phi_1,\ell_2,\Phi_2)=\bra{\ell_1,\Phi_1}\mathbf{1}_\text{phys}\ket{\ell_2,\Phi_2}=\begin{tikzpicture}[baseline={([yshift=-.5ex]current bounding box.center)}, scale=0.7]
    \pgftext{\includegraphics[scale=1]{sdworm1.pdf}} at (0,0);
    \draw (-2.9,-1) node {$\ell_1,\Phi_1$};
    \draw (2.9,-1) node {$\ell_2,\Phi_2$};
  \end{tikzpicture} \label{2.34worm}
\end{equation}
This can be achieved by first computing the norm on the co-invariant states $\ket{b}$, which we can compute using the KG inner product (or any other form of FP gauge-fixing) in the auxiliary Hilbert space $\ket{\ell,\Phi}$. 
This was explained very pedagogically in \cite{Held:2024rmg,Witten:2022xxp}, to which we refer readers for more orientation. We will compute the KG inner product at $\ell\to 0$. This simplifies the calculation, because we can use the plane wave form of the wavefunctions \eqref{2.24} in this regime. This results in
\begin{align}
    \braket{b_1|b_2}&=\int_{-\infty}^{+\infty}\d \ell \int_0^{2\pi}\d \Phi\,\Delta_\text{FP}\,\delta(\ell) \braket{b_1|\ell,\Phi}\braket{\ell,\Phi|b_2}\,,\quad \braket{\ell,\Phi|b}=Z_\text{trumpet}(b,\ell,\Phi)\nonumber\\&=-\frac{\i}{2} \int_0^{2\pi}\d \Phi \bigg(Z_\text{trumpet}(b_1,\ell\to 0,\Phi)^*\frac{\d}{\d \Phi}Z_\text{trumpet}(b_2,\ell\to 0,\Phi)\nonumber\\&\qquad\qquad\qquad\qquad\qquad\qquad\qquad-Z_\text{trumpet}(b_2,\ell\to 0,\Phi)\frac{\d}{\d \Phi}Z_\text{trumpet}(b_1,\ell\to 0,\Phi)^*\bigg)\nonumber\\&=\frac{1}{4\pi}\bigg(\frac{1}{b_1}+\frac{1}{b_2}\bigg)\int_0^{2\pi}\d \Phi\,e^{\i(b_2-b_1)\Phi}=\frac{1}{b_1}\delta_{b_1 b_2}\,.\label{2.35}
\end{align}
The states $\ket{b}$ span the physical Hilbert space (because they span the solutions of the WdW equation). The identity on this physical Hilbert space is therefore just the inverse of their inner product matrix\footnote{These states should be viewed as group averaged versions of states in the original Hilbert space, for the details see \cite{Held:2024rmg}.}
\begin{equation}
\mathbf{1}_\text{phys}=\sum_{b=1}^\infty b \ket{b}\bra{b}
\end{equation}
This leads to the double trumpet (or wormhole) amplitude
\begin{align}
     Z_\text{wormhole}(\ell_1,\Phi_1,\ell_2,\Phi_2)&=\left\langle \ell_1,\Phi_1\rvert\mathbf{1}_\text{phys} \rvert \ell_2,\Phi_2\right\rangle\nonumber\\&=\sum_{b=1}^\infty b \left\langle\ell_1,\Phi_1\rvert b\right\rangle \left\langle b \rvert \ell_2,\Phi_2\right\rangle =\sum_{b=1}^\infty b\, Z_b(\ell_1,\Phi_1) Z_b(\ell_2,\Phi_2)^*\label{2.35bis}
\end{align}
In the asymptotic limit \eqref{2.27} using the same holographic renormalization we obtain the double trumpet
\begin{equation}
    \boxed{Z_\text{wormhole}(\beta_1,\beta_2)=\sum_{b=1}^\infty b\,I_b(\beta_1/\hbar) I_b(\beta_2/\hbar)\,}\label{2.38}
\end{equation}
This is the universal answer for the connected expectation value in a finite cut Hermitian matrix integral of the partition function operator $\Tr e^{-\beta H/\hbar}$
\begin{equation}
    Z_\text{wormhole}(\beta_1,\beta_2)=\average{\Tr e^{-\beta_1 H/\hbar}\,\Tr e^{-\beta_2 H/\hbar}}_\text{conn}\text{at leading order}\label{2.39}
\end{equation}
This is the answer to leading order in $\dim(H)=2^N=\dim \mathcal{H}_\text{SYK}$. To derive equation \eqref{2.39}, one could rewrite the trumpet in the energy basis using Chebychev polynomials $T_b(-E)=\cos(b\arccos(-E))$
\begin{equation}
    Z_\text{trumpet}(b,\beta)=\int_{-1}^{+1}\d E\,e^{-\beta E/\hbar}\,\frac{1}{\pi}\frac{T_b(-E)}{\sqrt{1-E^2}}\,.
\end{equation}
Defining the spectral correlation $\rho_\text{wormhole}(E_1,E_2)$ via
\begin{equation}
    Z_\text{wormhole}(\beta_1,\beta_2)=\int_{-1}^{+1}\d E_1\,e^{-\beta_1 E_1/\hbar}\int_{-1}^{+1}\d E_2\,e^{-\beta_2 E_2/\hbar}\,\rho_\text{wormhole}(E_1,E_2)\,,\label{2.41}
\end{equation}
we find (combining equation \eqref{2.38} with equation (9.21) of \cite{Jafferis:2022wez})\footnote{We double checked this in the $\theta$ basis with $E=-\cos(\theta)$, where one can compute the finite cutoff wavefunctions either by Fourier transforming \eqref{2.18fctrumpet}, or by solving the WdW equation \eqref{2.6hwdw} using methods similar to those used in \cite{Goel:2020yxl}. The KG inner product of $\theta$ states is indeed  found (after a slightly more tedious calculation than \eqref{2.35}) to be the inverse of the wormhole amplitude in the $\theta$ basis.}
\begin{equation}
    \rho_\text{wormhole}(E_1,E_2)=\sum_{b=1}^\infty b\,\frac{1}{\pi}\frac{T_b(-E_1)}{\sqrt{1-E_1^2}}\frac{1}{\pi}\frac{T_b(-E_2)}{\sqrt{1-E_2^2}}=\frac{1}{2\pi^2}\frac{E_1E_2-1}{(E_1-E_2)^2}\frac{1}{\sqrt{1-E_1^2}}\frac{1}{\sqrt{1-E_2^2}}\,.\label{2.42}
\end{equation}
This is the universal spectral correlation of a finite cut (not double-scaled) matrix integral. For instance the same answer can be found in the ordinary Gaussian matrix integral, see for instance equations (2.34) and (2.35) in \cite{Blommaert:2021gha}. This is strong evidence that sine dilaton gravity is the path integral definition of the matrix integral whose leading order spectral density equals the exact spectrum of DSSYK
\begin{equation}
    \rho_\text{disk}(E)=2^N \sum_{b=-\infty}^{+\infty} \psi(b)\,\frac{1}{\pi}\frac{T_b(-E)}{\sqrt{1-E^2}}=\rho_\text{DSSYK}(E)\,.\label{2.43rho}
\end{equation}
This particular matrix integral was called q-deformed JT gravity \cite{Jafferis:2022wez,Jafferis:2022uhu}, and was claimed to be related with higher genus ``chord diagrams'' in \cite{Okuyama:2023byh}.

This is the first time that a gravity description is given of a \emph{finite cut} matrix integral with genuinely finite dimensional matrices. A priori there was no reason (except for chaotic universality \cite{Cotler:2016fpe}) to assume the wormhole amplitude in sine dilaton gravity would match with any matrix model.
A full proof would require matching more complicated observables such as higher genus amplitudes. We comment briefly on this in the discussion section \ref{sect:concl}.

We now show that equation \eqref{2.42} is actually the \emph{universal} spectral correlation in models of periodic dilaton gravity (introduced in \cite{Blommaert:2024whf}), which is strong evidence that these are all dual to finite cut matrix integrals. We consider the action
\begin{equation}
    I=\frac{1}{2}\int \d x \sqrt{g}\,(\Phi R+V(\Phi))+\int \d \tau\sqrt{h}\,\Phi K\,,\quad V(\Phi+2\pi)=V(\Phi)\,.
\end{equation}
We could think of these models as deformations of sine dilaton gravity by turning on $\lambda_b\sin(b\Phi)$, with $b$ integer. Expanding in $\lambda_b$ we get a gas of imaginary defects. These operators are periodic in $\Phi$ and thus physical. This should map to deformations of the matrix integral potential. Let us now solve the WdW equation for general periodic $V(\Phi)$. We consider the large class of potentials that results in a monotonic function $\theta(\Phi)$ for which the ADM energy is
\begin{equation}
    E=-\cos(\theta)\,.
\end{equation}
Here $V(\Phi)=2\,\d E/\d \Phi$ determines $E(\Phi)$. So $\theta=\arccos(-E(\Phi))$. We can then write the potential $V(\Phi)$ as
\begin{equation}
    V(\Phi)=2\,\frac{\d\theta}{\d \Phi}\sin(\theta)
\end{equation}
The WdW constraint equation \eqref{2.6hwdw} for general $V(\Phi)$ becomes
\begin{equation}
    H_\text{WdW}=\hbar^2\frac{\d}{\d \Phi}\frac{\d}{\d\ell}-\frac{1}{2}\ell\, V(\Phi)=\frac{\d\theta}{\d\Phi}\bigg(\hbar^2\frac{\d }{\d\theta}\frac{\d}{\d\ell}-\ell\,\sin(\theta)\bigg)=0\,.
\end{equation}
We can choose to study wavefunctions $\psi_b(\ell,\theta)$ with $\theta(\Phi)$. These satisfy the same differential equation as those of sine dilaton gravity \eqref{2.6hwdw}, resulting just in the replacement $\Phi\to \theta$ in \eqref{2.18fctrumpet}. Because $\theta$ is periodic too, \eqref{2.24} again implies that $b$ are integers.
In the KG inner product \eqref{2.35} we can transform to $\theta$ coordinates without Jacobian (this had to work, since $\theta(\Phi)$ is a targetspace diffeomorphism, and therefore redundant) so we find \eqref{2.35bis} with $\Phi\to \theta$. The asymptotic limit is $\theta=\pi/2+\i\infty$ such that we find exactly the same asymptotic trumpets \eqref{2.28}, which we may write as\footnote{Here we used that $\theta(\pi/2)=\pi/2$ for the class of periodic potentials $V(\Phi)$ studied in \cite{Blommaert:2024whf}. Also, if we suggestively write $E=W(\Phi)$ where $W'(\Phi)=V(\Phi)$, we can directly compare the integrand of this expression with that of the disk partition function.}
\begin{equation}
Z_\text{trumpet}(b,\beta)=\frac{1}{\pi}\int_{-1}^{+1}\frac{\d E}{\sqrt{1-E^2}} T_b(-E)\,e^{ -\beta E}\,.
\end{equation}
We may transform to $\Phi$ coordinates using $E(\Phi)$ but unlike for sine dilaton gravity this does not simplify the integral representation. It would be interesting to see if this expression can be found as well using the gas of defects approaches. One could generalize \eqref{eq:okudecomp} to generic periodic dilaton gravity using our proposal for on the exact disk partition function \cite{Blommaert:2024whf}, simply by Fourier transforming the disk density of states of these models as function of $\theta(\Phi)$ as in equation \eqref{2.43rho}.
We have thus shown that \eqref{2.38} and \eqref{2.42} hold for general periodic $V(\Phi)$.\footnote{Because $\theta\neq \Phi$ the state $b$ is not classically a simple defect anymore (since that would multiply $\Phi$ in the action, not $\theta$). But we could still expand these states in geodesic states (classically), because periodic functions in $\theta(\Phi)$ are periodic in $\Phi$.}

As a consistency check, notice that the same logic also applies to monotonic deformations of the JT gravity potential $2\Phi$. This bulk explanation for why the wormhole in double-scaled matrix integrals is universal circumvents the rather involved computation of the wormhole by summing a gas of defects.

\subsection{Comparing with analytically continued sinh dilaton gravity}\label{subsect:ungworm}
We briefly compare with what would have happened had we not taken the periodicity of $\Phi$ into account. The wavefunctions \eqref{2.18fctrumpet} are solutions in any case, and we can use the $\ell\to 0$ expansion \eqref{2.24}. Crucially, however, the $\Phi$ integral in the KG inner product \eqref{2.35} would range from $-\infty$ to $+\infty$, resulting in\footnote{We are using the nomenclature of \cite{Blommaert:2024whf} to refer to this setup as the ungauged theory.}
\begin{equation}
    \braket{b_1|b_2}_\text{ungauged}=\frac{1}{b_1}\delta(b_1-b_2)\,.
\end{equation}
This is indeed the inner product in sinh dilaton gravity or the Virasoro minimal string \cite{Collier:2023cyw}. It results in the same identity operator that also appears in the wormhole computation of JT gravity \cite{Saad:2019lba,Held:2024rmg}. This is the universal answer for double scaled matrix integrals (up to the choice of normalization of the wavefunctions, which affects the measure, but not the fact that this is diagonal)
\begin{equation}
    \mathbf{1}_\text{phys ungauged}=\int_0^\infty \d b  b \ket{b}\bra{b}\,.\label{2.49id}
\end{equation}
Thus, periodically identifying the dilaton takes one from a double-scaled matrix integral to a finite cut matrix integral, because the geodesic length $b$ of the wormholes gets discretized. 

We can track precisely how this occurs if we analytically continue from sinh gravity to sine gravity. The wormhole amplitude for sinh dilaton gravity has been computed in \cite{Collier:2023cyw}:
\begin{align}
    Z_\text{wormhole sinh}(\beta_1,\beta_2)&=\int_0^\infty \d \omega \omega\,\frac{1}{\pi} K_{\i \omega}(\beta_1/\hbar) \frac{1}{\pi}K_{\i \omega}(\beta_2/\hbar)\nonumber\\
    &=\int_0^\infty \d s_1\, e^{-\beta_1 \cosh(s_1)/\hbar}\int_0^\infty \d s_2\, e^{-\beta_2 \cosh(s_2)/\hbar} \int_0^\infty \d \omega \omega\,\frac{1}{\pi}\cos(\omega s_1) \frac{1}{\pi}\cos(\omega s_2)\nonumber\\&=\int_0^\infty \d E_1\,e^{-\beta_1 \text{cosh}(\sqrt{E_1})/\hbar}\int_0^\infty \d E_2\,e^{-\beta_2 \text{cosh}(\sqrt{E_2})/\hbar}\,\rho_\text{wormhole sinh}(E_1,E_2)\,.
\end{align}
This features the universal spectral correlation of double-scaled matrix integrals \cite{Saad:2019lba}
\begin{equation}
    \rho_\text{wormhole sinh}(E_1,E_2)=\int_0^\infty \d b b \,\frac{\text{cos}(b\sqrt{E_1})}{2\pi\sqrt{E_1}}\,\frac{\text{cos}(b\sqrt{E_2})}{2\pi\sqrt{E_2}}=-\frac{1}{4\pi^2}\frac{E_1+E_2}{(E_1-E_2)^2}\frac{1}{\sqrt{E_1}}\frac{1}{\sqrt{E_2}}\,.
\end{equation}
We may attempt continuation to ``ungauged'' sine dilaton gravity by replacing $s=\i \theta$ and $\i \omega=b$ resulting in\footnote{We also analytically continue the notions of $\beta$ and $\hbar$ in the usual way \cite{Blommaert:2024whf}.}
\begin{equation}
    Z_\text{wormhole ungauged}(\beta_1,\beta_2)=\int_{0}^{+\infty}\d \theta_1\,e^{\beta_1\cos(\theta_1)}\int_{0}^{+\infty}\d \theta_1\,e^{\beta_2\cos(\theta_2)}\int_0^\infty \d b b\,\frac{1}{\pi}\cos(b \theta_1)\frac{1}{\pi}\cos(b\theta_2)\,.\label{2.52}
\end{equation}
We may formally interpret this as the connected two-point correlator in a double-scaled matrix integral
\begin{equation}
    Z_\text{wormhole ungauged}(\beta_1,\beta_2)=\average{\Tr e^{\beta_1 \arccos(\sqrt{H})/\hbar}\,\Tr e^{\beta_1 \arccos(\sqrt{H})/\hbar}}_\text{conn}\text{at leading order}\,.\label{2.53}
\end{equation}
The continuous integrating measure $\d b b$ corresponds with the identity \eqref{2.49id}. \eqref{2.52} is exactly the same as the wormhole amplitude \eqref{2.41} in sine dilaton gravity up to an infinite overall prefactor. Indeed
\begin{align}
    \int_0^\infty \d \theta\,e^{\beta\cos(\theta)/\hbar}\,\frac{1}{\pi}\cos(b\theta)=\sum_{n=0}^\infty \delta(b=n) I_n(\beta/\hbar) \label{2.54}
\end{align}
Comparing with \eqref{2.38} leads then as to
\begin{equation}
    Z_\text{wormhole ungauged}(\beta_1,\beta_2)=\infty\,Z_\text{wormhole}(\beta_1,\beta_2)\,.
\end{equation}
The infinite prefactor accounts for the overcounting $\Phi(x)\to \Phi(x)+2\pi n$ in the path integral, which is an exact symmetry of the dilaton gravity path integral on the wormhole topology. So, one could obtain the wormhole amplitude \eqref{2.38} in sine dilaton gravity also by first considering an analytic continuation of sinh dilaton gravity \cite{Blommaert:2023wad,Mertens:2020hbs,Fan:2021bwt}, and then stripping off infinite prefactors from overcounting periodic shifts in the dilaton. This also works for the gravitational wavefunctions in the fully open channel \cite{Blommaert:2024whf}.

It is amusing that correlators of periodic operators such as \eqref{2.53} in a (non-periodic) double-scaled matrix integral are identical (up to overall infinities) to correlators in a finite cut matrix integral \eqref{2.39}. It would be interesting to derive such a reduction directly in the matrix theory by summing over images, projecting on the sector of the integrand invariant under the symmetry of the operator insertions.

\section{Boundary conditions and branes}\label{sect:classify}
In this section we classify possible boundary conditions of sine dilaton gravity. We find a two-parameter space of boundary conditions (and their Legendre duals) that correspond with FZZT brane boundary conditions in Liouville theory \cite{Fateev:2000ik}. Because sine dilaton gravity is a sum of two Liouville theories, there is indeed a two-parameter space of FZZT branes. In \textbf{section \ref{subsect:redundancies}} we show that there is a redundancy in description. The brane state is constrained by the WdW constraint, such that there is only one physical (invariant) FZZT brane parameter, which is canonically conjugate to the length $b$ of the trumpet \eqref{2.16hbarb}. In the remainder of the section we quantize sine dilaton gravity with these branes in three different channels. In \textbf{section \ref{subsect:3.3channels}} we detail these channels: closed, semi-open and fully-open. In \textbf{section \ref{subsect:3.4JTbranes}} we explain that in the effective AdS$_2$ metric \eqref{solutions}, our branes map to EOW branes \cite{Kourkoulou:2017zaj,Gao:2021uro} with fixed AdS$_2$ mass and curvature, and find the fully-open channel state corresponding with each pair of FZZT brane boundary conditions. In \textbf{section \ref{subsect:3.5semiopen}} we find the Hamiltonian generating time evolution in the semi-open channel (with Cauchy slices orthogonal to the brane), generalizing the JT calculations of \cite{Gao:2021uro}, and we show that the fully-open channel and the semi-open channel agree on the EOW brane disk amplitude. In \textbf{section \ref{subsect:examples}} we compare with branes introduced by Okuyama in \cite{Okuyama:2023byh}.

Asides from an a priori interest in FZZT branes, one reason for going through this exercise is that quantizing sine dilaton gravity in the presence of EOW branes allows for a derivation of the asymptotic trumpet amplitude, independent of the WdW quantization of section \ref{sect:trumpets}, by Fourier transforming the closed channel EOW brane amplitude. In \textbf{section \ref{subsect:4.4recover}} we show that this calculation reproduces \eqref{2.28}.

\subsection{Classifying boundary conditions and branes}
This subsection follows an approach similar to that for boundary conditions in JT gravity \cite{Goel:2020yxl}. Consider
\begin{align}
    I=\frac{1}{2}\int \d x \sqrt{g}\,(\Phi R+2\sin(\Phi))+\int \d u\sqrt{h}\,\Phi K-\overline{\mu} \int_\text{brane} \d u\sqrt{h}\,e^{\i \Phi/2}-\mu \int_\text{brane} \d u\sqrt{h}\,e^{-\i \Phi/2}\label{2.1}
\end{align}
We introduced the usual Liouville boundary cosmological constants \cite{Fateev:2000ik}. The cosmological constant $\mu$ couples to $L_\text{AdS}$ \eqref{1.26lengths} and $\overline{\mu}$ couples to $\overline{L}_\text{AdS}$. Variation of $I$ and using the bulk EOM leaves us with
\begin{equation}
    \delta I=\int_\text{brane}\d u \bigg\lbrace\,\delta \sqrt{h}\,\bigg(n^\mu \partial_\mu \Phi-\overline{\mu}e^{\i \Phi/2}-\mu e^{-\i \Phi/2}\bigg)+\sqrt{h}\, \delta \Phi\,\bigg(K-\frac{\i}{2}\overline{\mu}e^{\i \Phi/2}+\frac{\i}{2}\mu e^{-\i \Phi/2}  \bigg) \bigg\rbrace\,.\label{2.2bdyterms}
\end{equation}
The solutions of the bulk EOM are the usual black hole metrics \eqref{solutions}. One recognizes the two familiar Legendre conjugate pairs of dilaton gravity \cite{Goel:2020yxl}. One can either fix $\sqrt{h}$ or $n^\mu \partial_\mu \Phi$, and in addition we can either fix $\Phi$ or $K$. Using the boundary terms in \eqref{2.1}, the quantities that are being fixed are
\begin{equation}
    n^\mu \partial_\mu \Phi=\overline{\mu}e^{\i \Phi/2}+\mu e^{-\i \Phi/2}\,,\quad K=\frac{\i}{2}\overline{\mu}e^{\i \Phi/2}-\frac{\i}{2}\mu e^{-\i \Phi/2}\,.\label{2.3}
\end{equation}
This looks more intuitive upon introducing the extrinsic curvature $K_\text{AdS}$ in the AdS metric \eqref{solutions}, and a different extrinsic curvature $\overline{K}_\text{AdS}$ in the other Weyl-rescaled metric $e^{\i \Phi}\d s^2$. Introducing the Liouville fields $\varphi$ and $\overline{\varphi}$ as usual \cite{Blommaert:2024ydx}
\begin{equation}
    \rho-\i \Phi/2=\varphi\,,\quad \rho+\i \Phi/2=\bar{\varphi}\,,
\end{equation}
the on-shell variation of the action simplifies\footnote{Explicitly
\begin{equation}
    K_\text{AdS}=e^{\i \Phi/2}K-\frac{\i}{2}e^{\i \Phi/2}n^\mu \partial_\mu \Phi\,,\quad \overline{K}_\text{AdS}=e^{-\i \Phi/2}K+\frac{\i}{2}e^{-\i \Phi/2}n^\mu \partial_\mu \Phi\,.
\end{equation}
For comparison with the usual notation of FZZT boundary conditions \cite{Fateev:2000ik}, recall that $K_\text{AdS}=\partial_\mu n_\text{AdS}^\mu$ and $n_\text{AdS}^\mu=e^{-\varphi}\hat{n}^\mu$.}
\begin{equation}
    \delta I=\i \int_\text{brane}\d u \bigg\lbrace \bigg( \i\mu+K_\text{AdS}\bigg)\,\delta e^{\varphi}+\bigg( \i\overline{\mu}-\overline{K}_\text{AdS}\bigg)\,\delta e^{\overline{\varphi}} \bigg\rbrace 
\end{equation}
There are indeed the usual Neumann boundary conditions associated with FZZT branes \cite{Fateev:2000ik}:
\begin{equation}\label{eqn:Kvsmu}
    K_\text{AdS}=-\i \mu\,,\quad \bar{K}_\text{AdS}=\i \overline{\mu}\,.
\end{equation}
From $\delta I$ we also deduce the symplectic two-form on minisuperspace, which is intuitive given \eqref{2.1}:
\begin{equation}
    \omega=\d L_\text{AdS}\wedge \d K_\text{AdS}-\d \overline{L}_\text{AdS}\wedge \d \overline{K}_\text{AdS}\,.\label{2.9}
\end{equation}
Canonical quantization in the pre-Hilbert space (prior to imposing the WdW constraints) thus indeed proceeds via \eqref{2.13canstr}, so for instance $[L_\text{AdS},K_\text{AdS}]=\i\hbar$. Complex conjugation exchanges the boundary terms in the action \eqref{2.1}, so $\mu^*$ then couples to $e^{\overline{\varphi}}$. This plays a role in section \ref{subsect:3.5semiopen}. One could fix the lengths $L_\text{AdS}$ or $\overline{L}_\text{AdS}$ instead of the curvatures. The Legendre transform removes the boundary terms in the action \eqref{2.1}. We can think of the usual holographic boundary conditions \eqref{Irescaled} as corresponding with
\begin{equation}
    \mu=\i\,,\quad \overline{\mu}=-\i\cos(\theta)\,.\label{3.9assbranes}
\end{equation}
Indeed, then using \eqref{2.26ass} $\overline{L}_\text{AdS}=\i \beta$ one obtains $\i \cos(\theta)\overline{L}_\text{AdS}=\beta E$ which is the correct boundary term for fixed energy $E=-\cos(\theta)$ asymptotic boundary conditions. The last term in \eqref{2.1} is the holographic renormalization term in \eqref{Irescaled} provided indeed that $\mu=\i$.\footnote{Alternatively as discussed in section \ref{sect:trumpets} we may consider the Legendre conjugate boundary conditions $L_\text{AdS}=+\infty$. This is related with the close relation of ZZ brane boundary conditions and infinite lengths in Liouville theory \cite{Mertens:2020hbs}.}

\subsection{Redundancies of closed branes}\label{subsect:redundancies}
The two-parameter space of $(\mu,\overline{\mu})$-FZZT branes is redundant. Indeed, recall that a basis for the gauge-invariant Hilbert space of closed universes is spanned by the integer-length trumpets $\ket{b}$. We compute the single parameter conjugate to $b$ and the associated redundancies in the space $(\mu,\overline{\mu})$.

Recall the (classical) WdW constraint equation \eqref{2.24wdwh}:
\begin{align}\label{eqn:wdwconstraint}
    L_\text{AdS}^2(1-K_\text{AdS}^2) = \overline{L}_\text{AdS}^2(1-\overline{K}_\text{AdS}^2).
\end{align}
The length, $L_\text{AdS}$ of a closed $K_\text{AdS}$ curve is related to that of a homotopic geodesic curve by the equation
\begin{align}\label{eqn:3.37}
    \hbar b=L_0 = \sqrt{1-K_{\text{AdS}}^2} L_\text{AdS}.
\end{align}
The WdW Hamiltonian generates $\rho$-evolution in the metric \eqref{solutions} whilst conserving $L_0$. As the WdW Hamiltonian generates gauge transformations, we should read the constraint in \eqref{eqn:wdwconstraint} as saying that all closed branes which are homotopic to the same closed geodesic brane are gauge equivalent. Using this constraint and the symplectic form \eqref{2.9} for the closed brane phase space, we can derive the conjugate variable to the length $L_0$, for instance by eliminating $\overline{L}_\text{AdS}$. This results in
\begin{align}
    [b,\Phi_h]=\i\,,\quad \Phi_h=\text{arcsin}(K_\text{AdS}) - \text{arcsin}(\overline{K}_\text{AdS})\label{3.21}
\end{align}
Here $\Phi_h$ is the value of the dilaton at the horizon in the \emph{unique} AdS$_2$ metric \eqref{solutions} which has a Cauchy slice with curvature $K_\text{AdS}$ and dual curvature $\overline{K}_\text{AdS}$. In terms of FZZT brane parameters $(\mu,\overline{\mu})$, using equation \eqref{eqn:Kvsmu}, this determines the physical brane parameter to be
\begin{align}\label{eq:theta0}
    \boxed{\Phi_h= -\i\text{arcsinh}(\mu) - \i \text{arcsinh}(\overline{\mu})\,}
\end{align}
We can indeed check (using the operator ordering detailed in \eqref{2.24wdwh} that the combination $\text{arcsin}(K_\text{AdS}) - \text{arcsin}(\overline{K}_\text{AdS})$ commutes with $H_\text{WdW}$ and is thus a gauge-invariant quantity! This means that physical wavefunctions $\psi(K_\text{AdS},\overline{K}_\text{AdS})$ only depend on the invariant combination $\Phi_h(K_\text{AdS},\overline{K}_\text{AdS})$.\footnote{This argument mimics what happens for JT gravity in equation (4.1) of \cite{Held:2024rmg}.} Indeed, as key example we will see in section \ref{subsect:4.4recover} that the cylinder amplitude of sine dilaton gravity ending on a $(\mu,\overline{\mu})$ EOW brane \eqref{4.15EOWcylinder} only depends on the combination $\Phi_h(\mu,\overline{\mu})$
\begin{equation}
    \bra{\mu,\overline{\mu}}\mathbf{1}_\text{phys}\ket{\beta}=\sum_{b=0}^\infty \frac{e^{-\i b \Phi_h}}{1-q^{2b}}\,I_b(\beta/\hbar)=\begin{tikzpicture}[baseline={([yshift=-.5ex]current bounding box.center)}, scale=0.7]
    \pgftext{\includegraphics[scale=1]{sdworm3.pdf}} at (0,0);
    \draw (-2.2,0) node {\color{red}$\mu,\overline{\mu}$};
    \draw (1.7,-1) node {$\beta$};
  \end{tikzpicture}\label{3.14}
\end{equation}

So, closed $(\mu,\overline{\mu})$-branes with the same value of $\Phi_h$ are gauge equivalent. One can choose any member of the gauge orbit to represent a particular closed EOW brane. This is in particular convenient in section \ref{subsect:4.4recover} when we derive the trumpet  \eqref{2.28} by Fourier transforming EOW branes. We can choose the branes in the gauge orbit with the simplest semi-open Hamiltonian description (see section \ref{subsect:3.5semiopen}) to do that calculation in principle. Asymptotic fixed energy branes \eqref{3.9assbranes} have as invariant data $\Phi_h=\arccos(-E)$.

\subsection{Overview of quantization channels}\label{subsect:3.3channels}
Before presenting the canonical quantization of sine dilaton gravity with a $(\mu,\overline{\mu})$ EOW brane in section \ref{subsect:3.4JTbranes} and section \ref{subsect:3.5semiopen}, we present an overview of the different quantization channels we will use, and the results we aim to prove. We consider three channels.
\begin{enumerate}
    \item In the \textbf{fully-open channel}, we let the $(\mu,\overline{\mu})$ EOW brane coincide with a Cauchy slice, such that branes have interpretations as states $\ket{\mu,\overline{\mu}}$. The EOW brane disk amplitude is then computed as an overlap matrix element
    \begin{equation}
        \begin{tikzpicture}[baseline={([yshift=-.5ex]current bounding box.center)}, scale=0.7]
    \pgftext{\includegraphics[scale=1]{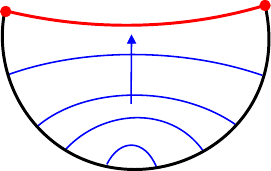}} at (0,0);
    \draw (0,1.6) node {\color{red}$\mu,\overline{\mu}$};
    \draw (1.8,-1.3) node {$\beta$};
    \draw (-1,0.1) node {\color{blue}time};
  \end{tikzpicture} \,\,\,=\bra{\mu,\overline{\mu}}e^{-\beta \mathbf{H}/\hbar}\ket{\mathbf{L}_\text{open}=0}\,.
    \end{equation}
    The Hamiltonian $\mathbf{H}$ describes time evolution on an open slice between two asymptotic boundaries, and is thus the ordinary sine dilaton gravity disk Hamiltonian \eqref{1.ham}.
    \item In the \textbf{semi-open channel}, the $(\mu,\overline{\mu})$ EOW brane lies orthogonal to Cauchy slices, which stretch from the EOW brane to the asymptotic boundary. This channel was the one used by Gao-Jafferis-Kolchmeyer to study EOW branes in JT gravity \cite{Gao:2021uro}. The Hamiltonian $\mathbf{H}_{\mu \overline{\mu}}$ which generates this flow depends on the specific brane under consideration. The EOW brane disk amplitude can thus alternatively be computed in this channel as the following transition amplitude:
    \begin{equation}
        \begin{tikzpicture}[baseline={([yshift=-.5ex]current bounding box.center)}, scale=0.7]
    \pgftext{\includegraphics[scale=1]{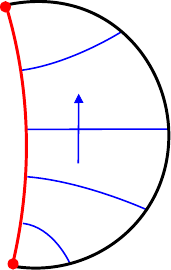}} at (0,0);
    \draw (-1.9,0) node {\color{red}$\mu,\overline{\mu}$};
    \draw (1.3,-1.8) node {$\beta$};
    \draw (0.65,-0.3) node {\color{blue}time};
  \end{tikzpicture}\,\, =\bra{\mathbf{L}_\text{semi}=0}e^{-\beta \mathbf{H}_{\mu \overline{\mu}}/\hbar}\ket{\mathbf{L}_\text{semi}=0}\,.\label{semi3.16}
    \end{equation}
    The non-trivial check that we present in section \ref{subsect:3.5semiopen} is that these two quantization channels agree on the explicit answer for the disk amplitude
    \begin{equation}
        \bra{\mu,\overline{\mu}}e^{-\beta \mathbf{H}/\hbar}\ket{\mathbf{L}_\text{open}=0}=\bra{\mathbf{L}_\text{semi}=0}e^{-\beta \mathbf{H}_{\mu \overline{\mu}}/\hbar}\ket{\mathbf{L}_\text{semi}=0}\,.\label{3.17EOWtocheck}
    \end{equation}
    \item In the \textbf{closed channel} used throughout section \ref{sect:trumpets} we consider a circular EOW brane and (gauged) time flow generated by $H_\text{WdW}$. In the semi-open channel, this is computed (following similar logic as in \cite{Gao:2021uro}) as the thermal trace
    \begin{equation}
        \begin{tikzpicture}[baseline={([yshift=-.5ex]current bounding box.center)}, scale=0.7]
    \pgftext{\includegraphics[scale=1]{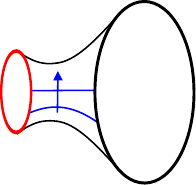}} at (0,0);
    \draw (-2.5,0) node {\color{red}$\mu,\overline{\mu}$};
    \draw (1.8,-1.3) node {$\beta$};
    \draw (-0.7,-1.2) node {\color{blue}time};
  \end{tikzpicture} =\Tr e^{-\beta \mathbf{H}_{\mu \overline{\mu}}/\hbar}
    \end{equation}
    The non-trivial check here is that the resulting amplitude \eqref{3.14} satisfies the closed channel WdW equation. This is clear from equation \eqref{3.14}, as the wavefunction in $b$ basis only depends on the gauge invariant conserved quantity $\Phi_h$.
\end{enumerate}
In the remainder of this section, we flesh out the fully-open and semi-open quantizations.

\subsection{Mapping to JT-branes and fully-open channel quantization}\label{subsect:3.4JTbranes}
We know that the boundary conditions \eqref{2.3} describe branes of fixed extrinsic curvature in an auxiliary AdS$_2$ metric. It may then not come as a surprise that at the classical level our branes can be identified with branes in an \emph{auxiliary} theory of JT gravity. In JT gravity, the natural set of boundary conditions are labeled by two parameters: the extrinsic curvature $K_\text{AdS}$ and the effective mass of the brane $m_\text{AdS}$, where the mass is related to the dilaton profile and extrinsic curvature via \cite{Gao:2021uro, Iliesiu:2020qvm}
\begin{align}
    m_\text{AdS} = \partial_{n_\text{AdS}}\Phi_\text{AdS} - \Phi_\text{AdS} K_\text{AdS}.\label{3.19}
\end{align}
We now show that the classical trajectories followed by $(\mu,\overline{\mu})$ branes in the effective AdS$_2$ are trajectories of branes in JT gravity with
\begin{align}\label{eqn:muGmuMvsKm}
    m_\text{AdS} = \overline{\mu} + \mu \cos(\theta)\,,\quad K_\text{AdS} = -i\mu.
\end{align}
Here $\theta=\Phi_h$ in sine dilaton gravity \eqref{1.9}. To check \eqref{eqn:muGmuMvsKm}, we insert in \eqref{3.19} the relation \eqref{1.11dilatons} between the dilaton and the AdS$_2$ radial coordinate $\Phi_\text{AdS}$
\begin{align}\label{eqn:phivsphijt}
    \Phi = \pi/2 + \i \log(\Phi_\text{AdS} +\i\cos(\theta))\,.
\end{align}
Using then \eqref{2.3} and the conformal transformation of the normal vector $n^\mu_\text{AdS}=e^{\i \Phi/2}n^\mu$ completes the proof. 

Equation \eqref{eqn:muGmuMvsKm} can be used to derive the fully-open channel state $\ket{\mu,\overline{\mu}}$ corresponding to the $(\mu,\overline{\mu})$ branes, which we discuss next. Time evolution in the fully-open channel is governed by the sine dilaton gravity Hamiltonian \cite{Blommaert:2024ydx} \eqref{1.ham}
\begin{equation}
    H=\cos(P_\text{open})\sqrt{1-e^{-L_\text{open}}}=\cos(\theta)\,.\label{3.22}
\end{equation}
The goal of this section is to find the wavefunction $\braket{\theta|\mu,\overline{\mu}}$. We do this by first computing $\braket{L_\text{open}|\mu,\overline{\mu}}$. To find that wavefunction we notice that the AdS mass associated with a fully-open slice can be written in two ways (we will soon explain how this is derived)
\begin{align}
    m_\text{AdS}&=\overline{\mu}+\mu \cos(P_\text{open})\sqrt{1-e^{-L_\text{open}}}\nonumber\\&=\i \sin(P_\text{open})\sqrt{1+\mu^2}\sqrt{1-e^{-L_\text{open}}}\,.\label{3.23}
\end{align}
Since $P_\text{open}=-\i\hbar\, \d/\d L_\text{open}$ acting on $\braket{L_\text{open}|\mu,\overline{\mu}}$, this gives a difference equation depending on $(\mu,\overline{\mu})$ that determines the wavefunction $\braket{L_\text{open}|\mu,\overline{\mu}}$. This Schr\"odinger equation simplifies classically to\footnote{For $L_\text{open}\to \infty$ this reduces to the closed channel equation \eqref{eq:theta0} since $P_\text{open}\to \Phi_h$ indeed in the large length limit \cite{Blommaert:2024whf}.}
\begin{equation}
    \overline{\mu}=\sinh(\i P_\text{open}-\text{arcsinh}(\mu))\sqrt{1-e^{-L_\text{open}}}\, .\label{3.24}
\end{equation}
The wavefunctions are more conveniently expressed using angular variables $(\alpha,\overline{\alpha})$ defined through
\begin{equation}
    \overline{\mu} = -\i \cos(\overline{\alpha})\,,\quad \mu = \i \cos(\alpha)\,.\label{3.25}
\end{equation}
Equation \eqref{3.24} is classical. Quantum mechanically, one natural operator ordering is\footnote{We do not have a bulk derivation of this operator ordering. Perhaps this could be provided by leveraging the underlying quantum group structure of sine dilaton gravity.}
\begin{align}\label{eqn:mugmumstates}
    \cos(\overline{\alpha})= \frac{1}{2}e^{\i\mathbf{P}_\text{open}+\i\alpha}\sqrt{1-e^{-L_\text{open}}}  + \frac{1}{2}\sqrt{1-e^{-L_\text{open}}}e^{-i\mathbf{P}_\text{open}-\i\alpha}\,.
\end{align}
We recognize the familiar recursion relation for q-Hermite polynomials with $L_\text{open}=\hbar n$:
\begin{align}\label{eqn:generalmunwf}
    \braket{n|\mu,\overline{\mu}}= \frac{H_n(\cos(\overline{\alpha})| q^2)}{\sqrt{(q^2;q^2)_n}}e^{-\i\alpha n}\,,
\end{align}
where the discretization of the length $L_{\text{open}}$ is achieved by gauging the $\mathbf{P}_{\text{open}} \to \mathbf{P}_{\text{open}} + 2\pi$ symmetry of \eqref{eqn:mugmumstates} \cite{Blommaert:2024whf}.
Energy eigenstates $\ket{\theta}$ satisfy this same differential equation for $\alpha=0$ and $\overline{\alpha}=\theta$ (see \eqref{1.ham} and \eqref{3.9assbranes}). The wavefunctions in the energy basis are therefore
\begin{equation}
    \braket{\theta|\mu,\overline{\mu}}=\sum_{n=0}^\infty \frac{e^{-\i \alpha n}}{(q^2,q^2)^n}H_n(\cos(\theta)|q^2)H_n(\cos(\overline{\alpha})|q^2)=\frac{(e^{-2\i\alpha};q^2)_\infty}{(e^{-\i\alpha}e^{\pm \i \theta\pm \i \overline{\alpha}};q^2)_\infty}\,.\label{eqn:FZZTwithasympboundary}
\end{equation}
The second equality is equation (5.7) of \cite{ISMAIL1987379} and it mimics the computation of the two-point function in DSSYK with $e^{\i\alpha}=e^{\Delta }$.\footnote{We want to think of $\Delta$ as a mass which appears in the particle action as $e^{-\Delta L_\text{open}/\hbar}$, so we rescaled the usual notion of $\Delta$ in DSSYK by $\hbar$.} Therefore the state $\ket{\mu,\overline{\mu}}$ corresponds in sine dilaton gravity with inserting a particle with label $\Delta$ and fixing the energy on the other side of the particle, with the relation
\begin{equation}
    \overline{\mu}=\i E\,,\quad \mu=\i\cosh(\Delta)\,.
\end{equation}
The first equation is quite obvious since we identify temperature as $\overline{L}_\text{AdS}=\i \beta$. The second is consistent with the usual fixed energy state which has $\Delta=0$ \eqref{3.9assbranes}. It seems similar to the usual relation between the label $\Delta$ of the representation of the quantum group SU$_q(1,1)$ which governs sine dilaton gravity \cite{Blommaert:2023opb}, and the associated Casimir $C(\Delta)=\cosh(\Delta)$, which indeed equals the sine dilaton gravity mass of the particle. It would be interesting to understand this relation better.

Before comparing these wavefunctions \eqref{eqn:FZZTwithasympboundary} with the results from semi-open quantization, we still need to explain how to derive equation \eqref{3.23} (a crucial equation). The first line is simple, we simply insert \eqref{3.22} into \eqref{eqn:muGmuMvsKm}. For the second line, we first use the fact that in AdS$_2$ the length $L_K$ of slices with fixed curvature $K_\text{AdS}$ is related with that of the homologous geodesic slice with length $L_\text{open}$ via equation \eqref{eqn:3.37}. The AdS$_2$ momentum $\Pi_K$ of this slice (in JT conjugate to $L_K$) is therefore also related to $\Pi_\text{open}$ by a simple rescaling. The AdS$_2$ mass $m_\text{AdS}$ is this conjugate to the length of the slice, up to a factor of $\i$. Therefore we have
\begin{align}
    -\i m_\text{AdS}= \Pi_\text{open} \sqrt{1+\mu^2}.\label{3.30}
\end{align}
The question is then how to express the JT momentum $\Pi_\text{open}$ of a geodesic slice as function of its length $L_\text{open}$ and momentum $P_\text{open}$. The JT momentum of a geodesic slice is the time derivative of its length, since the fully open Hamiltonian of \cite{Harlow:2018tqv} is quadratic in the momentum conjugate to $L_\text{open}$. Applying this to the length in sine dilaton gravity \cite{Blommaert:2024whf}
\begin{align}
    L_\text{open} = 2\log \frac{\cosh(\sqrt{1-H^2} T)}{\sqrt{1-H^2}}\,,
\end{align}
and eliminating $T$ (after taking the $T$ derivative to compute $\Pi_\text{open}$) results in
\begin{equation}
    \Pi_\text{open}=\sqrt{1-H^2-e^{-L_\text{open}}}.
\end{equation}
Combining this with $H(P_\text{open},L_\text{open})$ \eqref{3.22} gives the desired expression 
\begin{align}\label{eqn:momlength2}
    \Pi_\text{open}= \sin (P_\text{open}) \sqrt{1-e^{-L_\text{open}}}.
\end{align}
Finally, inserting this expression in \eqref{3.30} reproduces the second line of \eqref{3.23}. We stress that the JT momentum $\Pi_\text{open}$ (the $T$ derivative of the geodesic length), should not be confused with the canonical conjugate $P_\text{open}$ in sine dilaton gravity of $L_\text{open}$. They only coincide in the JT limit.\footnote{The Hamiltonian $H$ has an interpretation in terms of a q-deformed simple harmonic oscillator \cite{Okuyama:2023byh}, where $H$ is to the position variable and $L_\text{open}$ maps to the energy. Then $\Pi_\text{open}$ has the interpretation as the momentum conjugate to the position.}

\subsection{Brane Hamiltonians and semi-open channel quantization}\label{subsect:3.5semiopen}
In this section we quantize sine-dilaton gravity in the presence of $(\mu, \overline{\mu})$-branes in the semi-open channel \eqref{semi3.16}, where Cauchy slices have at one end an asymptotic boundary and at the other the $(\mu, \overline{\mu})$-brane. This is akin to what was done in \cite{Gao:2021uro} for JT gravity, where the slice ended on an asymptotic boundary and a brane with $K_\text{AdS} = 0$.

As in the previous section and in \cite{Blommaert:2024ydx}, yet again it is convenient to describe our quantization slices in the auxiliary AdS$_2$ metric. Just as in \cite{Gao:2021uro}, we consider the lengths $L_\text{semi}$ of semi-open geodesics ending orthogonally to the brane, and we describe the phase space of classical solutions in terms of the lengths of these slices. By extending the calculation of geodesic lengths $L_\text{semi}$ in AdS$_2$ of \cite{Gao:2021uro} to more general $(K_\text{AdS}, m_\text{AdS})$-branes, we find in appendix \ref{app:details} that the AdS length of geodesics ending on the branes is
\begin{align}\label{eqn:Ltsolution}
    L_\text{semi}= \log \frac{m_\text{Ads}+ \sqrt{(1-K_\text{AdS}^2) \Phi_{h\,\text{AdS}}^2 + \mjt^2} \cosh \left(\Phi_{h\,\text{AdS}} T \right)}{\sqrt{1-K_\text{AdS}^2} \Phi_{h\,\text{AdS}}^2}\,.
\end{align}
Here, $T$ is boundary time and $\Phi_{h\,\text{AdS}}$ is the AdS dilaton at the horizon. From the AdS metric solution \eqref{solutions} or from \eqref{3.21} we see that $\Phi_{h\,\text{AdS}}=\sin(\theta)=\sqrt{1-H^2}$. Indeed the ADM energy $H$ is not affected by the presence of the brane \cite{Gao:2021uro}. Inserting furthermore the expression \eqref{eqn:muGmuMvsKm} for $m_\text{AdS}(H,\mu,\overline{\mu})$ we find an expression $L_\text{semi}(T,H)$ which is also function of the non-dynamical parameters $(\mu,\overline{\mu})$. We may then search for the conjugate function $P_\text{semi}$ such that $\d T\wedge \d H=\d L_\text{semi}\wedge \d P_\text{semi}$ and invert this canonical transformation to find the Hamiltonian of sine dilaton gravity ending on an EOW brane $H(L_\text{semi},P_\text{semi})=H_{\mu \overline{\mu}}$. This procedure results in\footnote{One way to solve this instead of going through the tedious canonical transformations is to compute $\d L/\d T$ as a function of $L$ and $H$ using \eqref{eqn:Ltsolution}. We can now integrate $\d T/\d L$ to obtain $T(L,H)$. Then we ansatz $H = \cos(P) \sqrt{1-ae^{-L}-be^{-2L}} - ce^{-L}$ and demand the Hamilton equation $\d H/\d P (L,H) = \d L/\d T (L,H)$ is satisfied. This leads to a solution, which is of course unique (since classical mechanics is deterministic).}
\begin{align}\label{eqn:Hamgeneralmu}
   \boxed{H_{\mu \overline{\mu}} =\cos(P_\text{semi}) \sqrt{1-2\overline{\mu} e^{-L_\text{semi}} - e^{-2L_\text{semi}}} - \mu e^{-L_\text{semi}}\,}
\end{align}
To discuss quantization, as for the fully open channel \cite{Blommaert:2024ydx}, is is more convenient to bring this Hamiltonian in a non-Hermitian form by shifting $P_\text{semi}$ in such a way that the Hamiltonian becomes
\begin{equation}
    H_{\mu \overline{\mu}}=\cos(P_\text{semi})-\mu e^{-L_\text{semi}}-\frac12\big(e^{-2L_\text{semi}}+2\overline{\mu}e^{-L_\text{semi}}\big)\,e^{\i P_\text{semi}}.\label{3.38hh}
\end{equation}
Upon shifting $L_\text{semi}\to L_\text{semi}-i\overline{\alpha}-\i \pi/2$ this becomes (with a natural choice of operator ordering) the recursion relation for the Al-Salam-Chihara polynomials \cite{koekoek2010hypergeometric,Lenells:2021zxo,Jafferis:2022wez,Berkooz:2018jqr} $Q_n(\cos(\theta)|A,B,q^2)$ which solve\footnote{These polynomials are a generalization of the q-Hermite polynomials, and part of the even larger q-Askey scheme \cite{koekoek2010hypergeometric,Lenells:2021zxo}}
\begin{align}\label{eqn:ASCpoly}
    2\cos(\theta)= e^{-\i \mathbf{P}_\text{semi}} + \big(A+B\big)\, e^{-\mathbf{L}_\text{semi}} + \big(1-A Be^{-\mathbf{L}_\text{semi}}\big)\, e^{\i \mathbf{P}_\text{semi}} \big(1-e^{-\mathbf{L}_\text{semi}}\big)\,,
\end{align}
with $L_\text{semi}=\hbar n$. Indeed, one finds that \eqref{eqn:ASCpoly} matches with our Hamiltonian \eqref{3.38hh} provided that
\begin{align}\label{eqn:ABtheta}
    A = e^{\i\overline{\alpha}-\i\alpha}\,,\quad B = e^{\i\overline{\alpha}+\i\alpha}\,.
\end{align}
Al-Salam-Chihara polynomials are orthogonal with measure (see for instance equation (B.15) in \cite{Berkooz:2018jqr})\footnote{The square roots in the denominator arise in our wavefunctions because for non-Hermitian Hamiltonians left-and right eigenvectors differ \cite{Blommaert:2023opb,yao2018edge}. Here that means the left-and right eigenvectors differ by these roots in the denominator.}
\begin{equation}\label{eqn:ASCnormalization}
    \delta_{n_1 n_2}=\int_0^\pi \d\theta\,\frac{\big(AB,e^{\pm 2\i\theta};q^2\big)_\infty}{\big(Ae^{\pm \i \theta},Be^{\pm \i \theta};q^2)_\infty}\frac{Q_{n_1}(\cos(\theta_1)|q^2)}{\big(q^2,AB;q^2)_{n_1}^{1/2}} \frac{Q_{n_2}(\cos(\theta_1)|q^2)}{\big(q^2,AB;q^2)_{n_2}^{1/2}}\,.  
\end{equation}

With these wavefunctions we may compute the disk amplitude of sine dilaton gravity in the presence of an EOW brane. This is computed by the following transition matrix element
\begin{equation}
    \begin{tikzpicture}[baseline={([yshift=-.5ex]current bounding box.center)}, scale=0.7]
    \pgftext{\includegraphics[scale=1]{sdworm5.pdf}} at (0,0);
    \draw (-1.9,0) node {\color{red}$\mu,\overline{\mu}$};
    \draw (1.3,-1.8) node {$\beta$};
    \draw (0.65,-0.3) node {\color{blue}time};
  \end{tikzpicture}\,\,=\bra{\mathbf{L}_\text{semi}=0}e^{-\beta \mathbf{H}_{\mu \overline{\mu}}/\hbar}\ket{\mathbf{L}_\text{semi}=0}\,.
\end{equation}
For an explanation why the initial and final state must be chosen to be $L_\text{semi}=0$, we refer the reader to an in depth explanation for the (in this sense) similar case of the usual disk amplitude in the fully-open channel in \cite{Blommaert:2024whf}. Using the Al-Salam-Chihara polynomials which solve \eqref{eqn:ASCpoly}, we obtain
\begin{equation}
    \bra{\mathbf{L}_\text{semi}=0}e^{-\beta \mathbf{H}_{\mu \overline{\mu}}/\hbar}\ket{\mathbf{L}_\text{semi}=0}=\int_0^\pi \d \theta\,\big(e^{\pm 2\i \theta};q^2\big)_\infty \frac{\big(e^{2\i\overline{\alpha}};q^2\big)_\infty}{\big(e^{\i\overline{\alpha}}e^{\pm \i \alpha\pm\i \theta};q^2\big)_\infty}\,e^{\beta \cos(\theta)}.
\end{equation}
This compares favorably with the same calculation in the fully-open channel

\begin{equation}
        \begin{tikzpicture}[baseline={([yshift=-.5ex]current bounding box.center)}, scale=0.7]
    \pgftext{\includegraphics[scale=1]{sdworm4.pdf}} at (0,0);
    \draw (0,1.6) node {\color{red}$\mu,\overline{\mu}$};
    \draw (1.8,-1.3) node {$\beta$};
    \draw (-1,0.1) node {\color{blue}time};
  \end{tikzpicture}\,\,\, =\bra{\mu,\overline{\mu}}e^{-\beta \mathbf{H}/\hbar}\ket{\mathbf{L}_\text{open}=0}\,,\label{3.41}
    \end{equation}
provided that the fully-open channel wavefunctions are
\begin{equation}\label{eqn:mubmuthetaoverlap}
\braket{\mu,\overline{\mu}|\theta}=\frac{\big(e^{2\i\overline{\alpha}};q^2\big)_\infty}{\big(e^{\i\overline{\alpha}}e^{\pm \i \alpha\pm\i \theta};q^2\big)_\infty}\,.
\end{equation}
The reader may notice that this actually differs from the ket-wavefunction in \eqref{eqn:FZZTwithasympboundary} by swapping $\alpha\leftrightarrow\overline{\alpha}$. In terms of boundary cosmological constants \eqref{3.25} this stems from a swap \emph{and} complex conjugation
\begin{align}\label{eqn:mutrans}
    \mu \to \overline{\mu}^* \text{ and } \overline{\mu} \to \mu^*\,.
\end{align} 
This arises due to a complex conjugation of the full boundary action in the associated gravitational path integral \eqref{2.1} when we take a ket to a bra. Indeed complex conjugation exchanges the two Liouville fields $\d s\, e^{-\i \Phi/2}\leftrightarrow \d s\, e^{+\i \Phi/2}$. This swaps the meaning of the cosmological constants (and complex conjugates them, if they are complex numbers). We do not understand which principle determines the fact that we should be thinking about bras (as apposed to kets) in order for EOW branes in the fully-open channel to match with the semi-open channel analysis. We merely observe that apparently this is the correct procedure. It would be interesting to understand this better. 

In summary, the semi-open channel amplitude computation is akin to computing the gravitational path integral in the presence of certain boundary conditions set by the states in the fully-open amplitude. Accounting for the necessary complex conjugation \eqref{eqn:mutrans}, we have seen that open-channel amplitudes may be successfully extracted from the semi-open channel, and vice versa. This provides a non-trivial check on our formulae for both quantization channels. 

We emphasize that (just like the WdW Hamiltonian \eqref{2.6hwdw}, and the fully-open channel Hamiltonian \eqref{1.ham}), the Hamiltonian in the semi-open channel \eqref{eqn:ASCpoly} has an exact symmetry
\begin{equation}
    \mathbf{P}_\text{semi}\sim \mathbf{P}_\text{semi}+2\pi\,.\label{3.45}
\end{equation}
Much like previous symmetries, this symmetry is to be gauged in order to avoid unphysical divergences in the disk amplitude \eqref{3.41}\,.

\subsection{Okuyama branes}\label{subsect:examples}
Before we compute the trumpet via semi-open channel quantization in section \ref{subsect:4.4recover}, we zero in on a specific set of our branes that was actually discussed in the literature previously by Okuyama in \cite{Okuyama:2023byh}. Before the connection between chords and sine-dilaton gravity was established, Okuyama already studied a natural (albeit somewhat ad hoc) q-deformation of the Hamiltonian in the presence of an EOW brane in JT gravity \cite{Gao:2021uro}. The Hamilton equation he considered for EOW branes in DSSYK was\footnote{We thank Xuchen Cao for extended discussion, and for sharing some of his yet-to-be-published results on this subject.}
\begin{align}\label{eqn:okurec}
    2\cos(\theta)= e^{-\i \mathbf{P}_\text{semi}} + a e^{-\mathbf{L}_\text{semi}} + e^{\i \mathbf{P}_\text{semi}} \big(1-e^{-\mathbf{L}_\text{semi}}\big)\,.
\end{align}
Comparing with the recursion relation for the Al-Salam polynomials in \eqref{eqn:ASCpoly}, we see that the Okuyama branes correspond to $A = a$ and $B=0$. This can be achieved by $\alpha,\overline{\alpha}\to\i \infty$ with 
\begin{equation}
    -\mu/\overline{\mu}=a=e^{-\i \Phi_h} \quad \text{finite}.
\end{equation}
Here, we related Okuyama's parameter $a$ with the conjugate $\Phi_h$ of the trumpet length $b$ in the closed channel using equation \eqref{eq:theta0}. In this case the brane state wavefunction 
\eqref{eqn:mubmuthetaoverlap} becomes
\begin{equation}
\braket{\mu,\overline{\mu}|\theta}=\frac{1}{\big(e^{-\i\Phi_h\pm\i \theta};q^2\big)_\infty} = \sum_{n=0}^\infty e^{-\i\Phi_h n}\,\frac{H_n(\cos(\theta)| q^2)}{(q^2;q^2)_n}\,.
\end{equation}
In the interpretation of the q-deformed simple harmonic oscillator, this is the coherent state of the q-annihilation operator $e^{\i \mathbf{P}}\sqrt{1-e^{-L}}$ with eigenvalue $e^{\i \Phi_h}$. 
\section{Deriving trumpets from brane quantum mechanics}\label{subsect:4.4recover}
In section \ref{subsect:finitecutoff} we derived the trumpet amplitude $I_b(\beta)$ in periodic dilaton gravity by solving the WdW equation. Here we rederive this amplitude by Fourier transforming the brane systems of section \ref{subsect:3.5semiopen} in the semi-open channel. This is as a consistency check on our brane quantization and WdW discussions. 

Starting from knowledge of the EOW brane quantum mechanics we can compute the trumpet as a Fourier transform of the cylinder diagram\footnote{What we call $Z_\text{trumpet}(b,\beta)$ in this equation ends up differing by a Selberg factor $1/(1-q^{2b})$ from the answer we found in \eqref{2.28}. This would drop out in the double trumpet calculation as it would be simply absorbed in the $b$ gluing measure. Such a Selberg determinant also appeared in the analogous JT calculation \cite{Gao:2021uro}.}
\begin{align}
    Z_\text{trumpet}(b,\beta)&=\int_{\mathcal{C}}\d \Phi_h\,e^{\i b \Phi_h}\,\Tr e^{-\beta \mathbf{H}_{\mu\overline{\mu}}/\hbar}\,, \label{4.1zzzzzzz}
\end{align}
where $\mathcal{C}$ is the inverse Laplace transform contour. This follows from the discussion of section \ref{subsect:redundancies}, where we saw that the geodesic trumpet length $\hbar b$ \eqref{2.16hbarb} and $\Phi_h$ are canonical conjugates $[b,\Phi_h]=\i$. For JT gravity, such a calculation was previously indeed shown to recover the trumpet \cite{Gao:2021uro}. We claim that this calculation gives the same answer for branes with general $(\mu,\overline{\mu})$ as long as they have identical $\Phi_h(\mu,\overline{\mu})$. We leave the details of this technical calculation to appendix \ref{sec:trumpetdetails}. Let us just note here that the result is indeed only dependent of $\Phi_h$:
\begin{equation}
    Z_\text{EOW}(\Phi_h,\beta)=   \Tr e^{-\beta \mathbf{H}_{\mu\overline{\mu}}/\hbar} = \sum_{b=0}^\infty \frac{e^{-\i b \Phi_h(\mu, \bmu)}}{1-q^{2b}}\,I_b(\beta)=\begin{tikzpicture}[baseline={([yshift=-.5ex]current bounding box.center)}, scale=0.7]
    \pgftext{\includegraphics[scale=1]{sdworm3.pdf}} at (0,0);
    \draw (-2,0) node {\color{red}$\Phi_h$};
    \draw (1.7,-1) node {$\beta$};
  \end{tikzpicture} .\label{4.15EOWcylinder}
\end{equation}
This formula diverges due to the $b=0$ contribution much like the JT gravity result \cite{Gao:2021uro}. The wavefunction is the path integral of a particle that couples minimally to the AdS$_2$ metric (and so non-minimally to the sine dilaton gravity metric). Indeed, we recognize the Selberg determinant from excited single-particle states on a hyperbolic cylinder with size $\hbar b$. Equation (9.52) of \cite{Jafferis:2022wez} suggests that the QFT path integral might not have a $b\to 0$ divergence, but leads to Hagedorn behavior. It would be interesting to understand this from a sine dilaton gravity point of view, perhaps by understanding better the quantum group isometries of the theory \cite{Blommaert:2023opb,Berkooz:2022mfk,Lin:2023trc,Almheiri:2024ayc}.

We stress that Okuyama also obtained equation \eqref{4.15EOWcylinder} by Fourier transforming his branes \cite{Okuyama:2023byh}. His impressive calculation uses several special formulas for sums of q-Hermite polynomials. Our contribution detailed in appendix \ref{sec:trumpetdetails} is a calculation which generalizes to all Al-Salam-Chihara polynomials defined by the recursion relation in \eqref{eqn:ASCpoly}. This calculation further explains the mechanism of brane redundancy. If one takes the discussion about brane equivalences in the closed channel in section \ref{subsect:redundancies} as a fact, then the main point of this section is redundant. In particular, with the knowledge of brane redundancy, we may simply gauge-fix to the Okuyama branes and import his result \cite{Okuyama:2023byh} for the Fourier transform.

We note that gauging the symmetry \eqref{3.45} already implies discretization of $b$ in the closed channel. Indeed, $\Tr e^{-\beta \mathbf{H}_{\mu \bmu}/\hbar}$ as an explicit function of $\Phi_h$ is invariant under $\Phi_h\to \Phi_h+2\pi$. In principle one could try a Fourier transform \eqref{4.1zzzzzzz} with $-\infty<\Phi_h<+\infty$. Because of the periodicity of the integrand this localizes on integer $b$ just like equation \eqref{2.54}.

\subsection{Semiclassical check}\label{subsect:4.3semi}

Since the exact derivation presented in appendix \ref{sec:trumpetdetails} of the trumpet wavefunction from a trace in the semi-open channel was quite technical, we use the remainder of this section to perform a simple semiclassical check of this result. We will compute the on-shell action of the semi-open brane quantum mechanics \eqref{3.38hh}, and compare this with the classical gravity calculation. 

In sine-dilaton gravity, the geodesic trumpet boundary corresponds with inserting a complex conical defect \eqref{2.23}
\begin{equation}
     \begin{tikzpicture}[baseline={([yshift=-.5ex]current bounding box.center)}, scale=0.7]
    \pgftext{\includegraphics[scale=1]{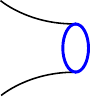}} at (0,0);
    \draw (1.15,0) node {\color{blue}$b$};
    \draw (-1,0) node {$\dots$};
  \end{tikzpicture}\,\,\Leftrightarrow\,\,\,V_b=\int \d x \sqrt{g}\,e^{(\i b-2\pi/\hbar)\Phi}\,.
\end{equation}
The computation of the on-shell action for dilaton gravity with such a defect was explained leading up to equation (4.34) of \cite{Blommaert:2024ydx}, following \cite{Dong:2022ilf,Blommaert:2023vbz}. The answer is
\begin{equation}
    I=\i b \Phi_h+\frac{1}{\hbar}\beta \cos(\Phi_h)\,.\label{4.17}
\end{equation}
Here $\Phi_h$ is the dilaton at the horizon, representing the lower-dimensional equivalent of the area $A = \Phi_h$. This is an integration variable in the gravity calculation. We note that this is consistent with our exact trumpet partition function \eqref{2.28}, which can be written as
\begin{equation}
    Z_\text{trumpet}(b,\beta)=I_b(\beta/\hbar)=\frac{1}{2\pi}\int_0^{2\pi}\d \Phi_h\,e^{\i b \Phi_h+\beta \cos(\Phi_h)/\hbar}\,.
\end{equation}
The saddle of this $\Phi_h$ integral reproduces the correct relation between the periodicity of the time circle and the length of the geodesic in the AdS$_2$ metric \eqref{solutions}, namely $\sin(\Phi_h)\beta=\i \hbar b$.

We now aim to reproduce \eqref{4.17} using brane quantum mechanics. Following equation \eqref{4.1zzzzzzz}
\begin{equation}
    Z_\text{trumpet}(b,\beta)=\frac{1}{2\pi}\int_0^{2\pi}\d \Phi_h\,e^{\i b \Phi_h}\,\Tr e^{-\beta \mathbf{H}_{\mu \overline{\mu}}/\hbar}\label{4.19}
\end{equation}
The path integral for the brane quantum mechanics is
\begin{align}
    \Tr e^{-\beta \mathbf{H}_{\mu \overline{\mu}}/\hbar}=\int \mathcal{D}P_\text{semi}\mathcal{D}L_\text{semi}&\exp\bigg(\frac{1}{\hbar}\int_0^\beta\d \tau \bigg\{\i P_\text{semi}\frac{\d}{\d \tau}L_\text{semi}\nonumber\\&\qquad +\cos(P_\text{semi})-\mu e^{-L_\text{semi}}-\frac{1}{2}\big(e^{-2L_\text{semi}}+2\overline{\mu}e^{-L_\text{semi}}\big)\,e^{\i P_\text{semi}}\bigg\}\bigg)
\end{align}
Because we compute the trace, we must look for periodic solutions, and the geometric picture of a $U(1)$ symmetric cylinder motives us to look for time-independent ones. Assuming $\d P_\text{semi}/\d \tau=\d L_\text{semi}/\d\tau=0$ the equations of motion read
\begin{equation}
   e^{\i P_\text{semi}} = -\frac{\mu}{\overline{\mu} + e^{-L_\text{semi}}}\,, \qquad e^{-2L_\text{semi}}+ 2\overline{\mu}e^{-L_\text{semi}}+\frac{\overline{\mu}^2-\mu^2}{1+\mu^2}=0\,.
\end{equation}
Solving the quadratic equation in $e^{-L_\text{semi}}$ and plugging the solution back into the one for $e^{\i P_\text{semi}}$ gives

\begin{equation}
    e^{-L}= -\bar\mu - \mu \,\sqrt{\frac{1+\bar{\mu}^2}{1+\mu^2}}\,, \quad e^{\i P_\text{semi}} = \sqrt{\frac{1+\mu^2}{1+\bar{\mu}^2}}\,.
\end{equation}
The on-shell action of this static solution is only its Hamiltonian, which evaluates on-shell to
\begin{equation}
    H_{\mu \overline{\mu}}=-\cosh(\text{arcsinh}(\mu)+\text{arcsinh}(\overline{\mu}))=-\cos(\Phi_h)\label{4.23}
\end{equation}
Combining with the Fourier transform in \eqref{4.19}, we recover the gravitational on-shell action \eqref{4.17} from brane quantum mechanics, including the fact that $\Phi_h$ is an integration parameter. This completes our check. The fact that this on-shell action of the quantum mechanics \eqref{4.23} for the $(\mu,\overline{\mu})$-brane quantum mechanics depends only on the combination $\Phi_h$ identified in \eqref{eq:theta0} is the semiclassical remnant of the redundancy in the closed Hilbert space. This pre-Hilbert space $L^2(\mu)\otimes L^2(\overline{\mu})$ reduces to a one-parameter physical Hilbert space $L^2(\Phi_h)$ with $\Phi_h\sim \Phi_h+2\pi$ (or square summables on integers $b$).

\section{Matrix integral dictionary}\label{sect:7matrixdictionary}
It might be useful to write down the dictionary between geodesic boundaries and FZZT branes (closely related with the EOW branes studied in the previous sections as we will soon see) in sine dilaton gravity, and certain operators in the (finite cut) matrix integral dual. This section should be read with the caveat in mind that we have not yet provided an all-genus definition of sine dilaton gravity that actually reproduces the matrix integral.

Geodesic boundaries correspond with\footnote{Technically we should subtract the one point function $2 \psi(b)/b$ \eqref{2.32} but we leave this implicit. We used equation \eqref{2.43rho} for the one point function in energy basis and the factor of 2 comes from the two signs of $b$.}
\begin{equation}
    \begin{tikzpicture}[baseline={([yshift=-.5ex]current bounding box.center)}, scale=0.7]
    \pgftext{\includegraphics[scale=1]{sdworm7.pdf}} at (0,0);
    \draw (1.15,0) node {\color{blue}$b$};
    \draw (-1,0) node {$\dots$};
  \end{tikzpicture}\,\,\Leftrightarrow\,\,\,\mathcal{O}_\text{G}(b)=\frac{2}{b}\Tr \cos( b \arccos(-H))\,.\label{7.1OG}
\end{equation}
This can be derived by imposing that
\begin{equation}
     \begin{tikzpicture}[baseline={([yshift=-.5ex]current bounding box.center)}, scale=0.7]
    \pgftext{\includegraphics[scale=1]{sdworm2.pdf}} at (0,0);
    \draw (-1.55,0) node {\color{blue}$b$};
    \draw (1.4,-1) node {$\beta$};
  \end{tikzpicture}=\average{\mathcal{O}_\text{G}(b) \Tr e^{-\beta H/\hbar}}_\text{conn}=Z_\text{trumpet}(b,\beta)\,.
\end{equation}
We can check this by inserting the double trumpet \eqref{2.42} for the connected correlator in energy basis, writing it as a sum of Chebychev polynomials and using their orthogonality relation
\begin{equation}
    \int_{-1}^{+1}\frac{\d E}{\sqrt{1-E^2}}T_{b_1}(-E) T_{b_2}(-E)=\frac{\pi}{2}\delta_{b_1 b_2}\,.
\end{equation}
Using the same steps (orthogonality twice) this also leads to the correct inner product a.k.a wormhole, see equation \eqref{2.35}
\begin{equation}
    \braket{b_1|b_2}=\average{\mathcal{O}_\text{G}(b_1) \mathcal{O}_\text{G}(b_2)}_\text{conn}=\frac{1}{b_1}\delta_{b_1b_2}\,.
\end{equation}

We learned from the calculation in section \ref{subsect:4.4recover} that EOW branes with physical parameter $\Phi_h$ expand in geodesic boundaries with wavefunction that we can read off from \eqref{4.15EOWcylinder}

\begin{equation}
    \mathcal{O}_\text{EOW}(\Phi_h)=\sum_{b=1}^\infty \frac{e^{-\i b \Phi_h}}{1-q^{2b}}\,\mathcal{O}_\text{G}(b)\,.
\end{equation}
In the matrix integral, FZZT branes are more interesting objects than EOW branes. They are closely related, but not identical. FZZT branes expand into geodesic boundaries as follows:
\begin{equation}
    \mathcal{O}_\text{FZZT}(\Phi_h)=-\sum_{b=1}^\infty e^{-\i b \Phi_h}\,\mathcal{O}_\text{G}(b)\,.\label{7.6fzzt}
\end{equation}
In JT gravity $\mathcal{O}_\text{G}(b)=2\Tr\text{cos}(b\sqrt{H})/b$ \cite{Blommaert:2021fob} and one obtains the relation $\mathcal{O}_\text{FZZT}(\Phi_h)=\Tr\log (H-\Phi_h^2)$ \cite{Saad:2019lba,Ponsot:2001ng,Hosomichi:2008th,Maldacena:2004sn}. EOW branes are classically identical to FZZT branes but they have a Selberg determinant included (from particle trajectory fluctuations). In sine dilaton gravity, the sum of particle states leads to the following relation between brane states:
\begin{equation}
   \mathcal{O}_\text{EOW}(\Phi_h)=-\sum_{n=0}^\infty \mathcal{O}_\text{FZZT}(\Phi_h+\i n \hbar)\,.
\end{equation}
The equation for JT is identical with the replacement $\hbar\to 1$, and explicitly doing the sum over $n$ gives the correct EOW brane dictionary $\mathcal{O}_\text{EOW}(\Phi_h)=\Tr \text{log} (\Phi_h\pm \sqrt{H})$ \cite{Gao:2021uro}. Summing over $b$ in \eqref{7.6fzzt} results in\footnote{We are suppressing $H$ independent additive constants that can be interpreted as one point functions much like in \eqref{7.1OG}.}
\begin{equation}
    \boxed{\mathcal{O}_\text{FZZT}(\Phi_h)=\Tr \log (H+\cos(\Phi_h))\,}\label{d.39}
\end{equation}
This is the natural replacement for $\Tr\log(H-\Phi_h^2)$ in finite-cut matrix models where $\Phi_h$ is an angular variable. Indeed for JT $E(\Phi_h)=\Phi_h^2$ whereas for sine dilaton gravity $E(\Phi_h)=-\cos(\Phi_h)$. As consistency check, we can solve the WdW equation \eqref{2.24wdwh} in the conjugate variables \eqref{2.13canstr}. In the asymptotic gauge \eqref{3.9assbranes} where $K_\text{AdS}=1$ and $\overline{K}_\text{AdS}=\cos(\Phi_h)$ one finds the following solution to the WdW equation
\begin{equation}
    \psi_b(\Phi_h)=-\frac{\i}{\sin(\Phi_h)}e^{-\i b \Phi_h}\,.\label{7.9wave}
\end{equation}
This matches as expected with the resolvent operator in the matrix integral
\begin{equation}
    \begin{tikzpicture}[baseline={([yshift=-.5ex]current bounding box.center)}, scale=0.7]
    \pgftext{\includegraphics[scale=1]{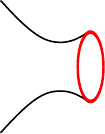}} at (0,0);
    \draw (1.45,0) node {\color{red}$\Phi_h$};
    \draw (-1,0) node {$\dots$};
  \end{tikzpicture}\,\,\Leftrightarrow\,\,\, \Tr \frac{1}{H+\cos(\Phi_h)}=\sum_b b\, \psi_b(\Phi_h)\,\mathcal{O}_\text{G}(b)
\end{equation}
Indeed, one can get this resolvent operator by taking a $\cos(\Phi_h)$ derivative of the FZZT boundary state, bringing down $\i b/\sin(\Phi_h)$ and thus reproducing the wavefunction \eqref{7.9wave}. Of course if we inverse Laplace transform this we recover $\Tr e^{-\beta H/\hbar}$.

\section{Concluding remarks}\label{sect:concl}
We discussed the canonical quantization of sine dilaton gravity on closed slices using WdW quantization, and the quantization of sine dilaton gravity in the presence of open and of closed EOW branes. Together with our quantization of the disk \cite{Blommaert:2024whf}, we presented a quite complete picture of the Hamiltonian formulation of sine dilaton gravity. In all channels, the periodicity of the dilaton implies certain discrete shift symmetries in the Hamiltonian, which are to be gauged, in order to avoid divergences. These shift symmetries discretize the lengths of spacetime. They are responsible for a UV complete-type spectrum \eqref{1.2pic}.

Whilst the existence of these symmetries is undeniable in the classical phase space, and our choice of operator ordering preserves this symmetry exactly quantum mechanically, it is not obvious to us what these symmetries correspond to at the path integral level. This is reflected in the fact that we do not yet know how to precisely quantize the theory on higher genus surfaces.\footnote{One reasonable definition of sine dilaton gravity on generic topologies was presented in \cite{Collier:2025pbm}. As explained in section 7.2 of \cite{Blommaert:2024whf} this definition differs significantly from ours already at disk level. One could trace this distinction to how one treats the different symmetries when quantizing: $b_{\text{CFT}} \to 1/b_{\text{CFT}}$ (broken for us, preserved for them) versus the discrete shift symmetry $\textbf{P} \to \textbf{P} + 2\pi$ (preserved and gauged for us, broken for them). Importantly for our purposes, their disk does not have the features which we seek to understand by studying sine dilaton gravity, see the introduction section \ref{sect:intro}. } 

We end this paper with several concluding remarks aimed towards future investigations. In \textbf{section \ref{subsect:d1matrix}} we discuss some potentially interesting future directions associated with the matrix integral duality.  In \textbf{section \ref{subsect:82generalizedminimal}} we suggest a way to define our theory on higher genus topologies, inspired by generalized minimal models \cite{Zamolodchikov:2005fy,Dotsenko:1984ad,Belavin:2005yj}. In \textbf{section \ref{subsect:nonoboundary}} we make several speculative comments on the relation between the Gaussian wavefunction \eqref{2.32} which replaces the no-boundary state in sine dilaton gravity, and the addition of observer degrees of freedom in quantum gravity.

\subsection{Matrix inspiration} \label{subsect:d1matrix}
Our wormhole amplitude \eqref{2.38} is consistent with a finite cut matrix integral. Finite cut matrix models admit a topological recursion which was worked out in \cite{10.1093/qmath/has004}. Genus $g$ amplitudes decompose universally as\footnote{A finite cut matrix integral is a matrix model with the property that if one computes the saddle point of the action we find a density of eigenvalues $\rho(E)$ with positive support on a finite interval.}
\begin{equation}\label{decomposition}
Z_{g,n}\left(\beta_1\dots \beta_n\right)=\sum_{b_1 \dots b_n=1}^{\infty} V_{g,n} \left(b_1\dots b_n\right) \prod_{i=1}^{n} b_i\, Z_\text{trumpet}(b_i,\beta_i)\,,
\end{equation}
with $Z_\text{trumpet}(b,\beta)=I_b(\beta/\hbar)$, and $V_{g,n}(b_1\dots b_2)$ polynomials (up to certain selection rule factors) in $b_i^2$. Universally
\begin{equation}
    V_{0,2}(b_1,b_2)=\frac{1}{b_1}\delta_{b_1b_2}\,.
\end{equation}
We recovered this via WdW canonical quantization as the universal answer for the wormhole amplitude in periodic dilaton gravity in section \ref{sect:doubletrumpet}. So this indeed makes an identification with finite cut matrix models possible at this point, even though it is certainly not guaranteed. This opens up several potential future directions. We list a few here.
\begin{enumerate}
    \item \textbf{All genus sine dilaton gravity.} It would be interesting to establish an all-genus duality between sine dilaton gravity and the finite cut q-deformed JT gravity matrix model \cite{Jafferis:2022wez,Okuyama:2023kdo}. The easiest amplitude to match is the three holed sphere (we suppress some prefactors)
    \begin{equation}
        V_{0,3}(b_1,b_2,b_3)=\frac{1+(-1)^{b_1+b_2+b_3}}{2}\,.\label{6.3volum}
    \end{equation}
    Another relatively simple amplitude is the genus one partition function
    \begin{equation}
        V_{1,1}(b)=\bigg(\frac{b^2-4}{48}+\frac{1}{8}\sum_{n=1}^\infty \frac{1}{\sinh(\hbar n/2)^2}\bigg) \frac{1+(-1)^{b}}{2}\,.
    \end{equation}
    The three holed sphere looks possibly consistent with the Liouville structure constants associated with our trumpet operator insertions \eqref{2.23}, and suggests a relation of our theory with generalized minimal models \cite{Zamolodchikov:2005fy,Dotsenko:1984ad}, which indeed have discrete (but infinite) operator spectra. We speculate on a relation with generalized minimal models in section \ref{subsect:82generalizedminimal}.

        \item \textbf{Exact quantization periodic dilaton gravity.} If the full matrix integral correspondence were established, it should be feasible to find an exact quantization of general periodic dilaton gravity \cite{Blommaert:2024whf}. One could do the gas-of-defects expansion \cite{Maxfield:2020ale,Witten:2020wvy} for periodic deformations away from sine dilaton gravity. It should then be possible to use e.g. intersection theory techniques (which indeed exist for finite cut matrix models \cite{10.1093/qmath/has004})  to resum the defects, and compute for instance the exact genus zero spectral density. This could also be studied directly in matrix integral parlance, the exponential of defects is a D-brane (or determinant operator) which modifies the matrix potential in a computable way. It is a one-line calculation to derive the associated spectrum. Good evidence for the exact matrix integral duality would be if one could recover the semiclassical spectrum of periodic dilaton gravity by expanding into defects \cite{Kruthoff:2024gxc}, writing those as matrix integral D-branes and seeing if the change in matrix potential reproduces the correct spectrum.\footnote{In \cite{Witten:2020ert,Witten:2020wvy} this was studied for JT-type (double-scaled) dilaton gravity.}

    \item \textbf{Relations with finite cutoff JT.} There are several ways in which our canonical quantization of sine dilaton gravity surprisingly resembles JT gravity at finite cutoff \cite{Iliesiu:2020zld}. Finite cutoff JT has a finite-support spectrum too, and the wormhole amplitude \cite{Griguolo:2021wgy} closely resembles our sine-dilaton answer \eqref{2.38}
    \begin{equation}
            Z_{\text{JT finite}}(\beta_1,\beta_2)=\frac{\beta_1 \beta_2}{\epsilon^2(\beta_{1}^2-\beta_{2}^{2})} (\beta_1 I_{0}(\beta_2/\epsilon^2)I_{1}(\beta_1/\epsilon^2)-\beta_2 I_{0}(\beta_1/\epsilon^2)I_{1}(\beta_2/\epsilon^2))\,,
    \end{equation}
    with the identification $\hbar=1/\varepsilon^2$ which matches the respective ends of the two spectra.\footnote{Our wormhole amplitude \eqref{2.38} indeed looks somewhat similar
    \begin{equation}
        \frac{\beta_1 \beta_2}{2 \hbar (\beta_{1}^2-\beta_{2}^{2})} (\beta_1 I_{0}(\beta_1/\hbar)I_{1}(\beta_2/\hbar)-\beta_2 I_{0}(\beta_1/\hbar)I_{1}(\beta_2/\hbar)+\beta_1 I_{1}(\beta_1/\hbar)I_{0}(\beta_2/\hbar)-\beta_2 I_{1}(\beta_1/\hbar)I_{0}(\beta_2/\hbar))\,.
    \end{equation}
    } It would be interesting to understand if finite cutoff JT could be given a genuine finite-cut matrix integral interpretation.

    Another relation with finite cutoff AdS is the following. As we have shown, there are two possible ways to map the classical solutions of sine-dilaton gravity to an AdS$_2$ disk. If we conformally map the Euclidean spacetimes to the full (asymptotic) disk in one AdS (by taking $K_{\text{AdS}} = 1$), then in the $\overline{\text{AdS}}$-spacetime the boundary lies at a finite distance. In particular, at fixed boundary inverse temperature $\beta$, we have $\overline{L}_{\text{AdS}} = \i \beta$,  where $\overline{L}_{\text{AdS}}$ is the \emph{non}-renormalized length of the boundary in $\overline{\text{AdS}}$. So we have a finite cutoff disk present.
\end{enumerate}
In the topological limit \cite{Blommaert:2024whf,Okuyama:2023kdo}, $V_{g,n}(b_1\dots b_n)$ can be interpreted as counting graphs with discrete lengths. These are the graph duals of the Wick contractions in the Gaussian matrix integral \cite{Gopakumar:2011ev}. This suggest a relation between chords (the discreteness of length in sine dilaton gravity) and the discrete laminations in finite cut matrix models, an intuitive concept explained in \cite{Gopakumar:2011ev} (for the interested reader).
\subsection{Generalized minimal models, screening charges and periodic topology change}\label{subsect:82generalizedminimal}
To advance towards a formulation of sine dilaton gravity on higher topologies, we consider the physical spectrum of operators in the Liouville formulation. We will follow the conventions of section 5 of \cite{Blommaert:2024ydx}.\footnote{Upon $\i \chi\to \chi$ these become the conventions of \cite{Mertens:2020hbs}.} We view the $\chi$ field as matter. Let us consider the bulk vertex operators in the double Liouville CFT, which is achieved by gravitational dressing of matter operators $\mathcal{O}_{\alpha_\text{M}}=\exp \left(2 \i \alpha_{\text{M}} \chi\right)$ as:
\begin{equation}
    V_{\alpha_\text{M}}=\int_{\Sigma} \mathcal{O}_{\alpha_M} \exp \left(2 \alpha \phi \right) \d x=\int_{\Sigma} \exp \left(2 \rho (\alpha+\alpha_\text{M})/b_\text{CFT}\right) \exp \left(\i \Phi (\alpha-\alpha_\text{M})/b_\text{CFT} \right) \d x\,.
\end{equation}
These are diff invariant i.e.\,$(1,1)$ primary operators if $\alpha=b_\text{CFT}-\alpha_M$. Consider for the sake of argument the matter degenerate operators
which are labeled by two positive integers $\left(m,n\right)$ with
\begin{equation}\label{alpha_deg}
\alpha_{\text{M}}=-(n-1)\frac{b_\text{CFT}}{2}+(m-1)\frac{1}{2b_\text{CFT}}\,.
\end{equation}
These correspond with the following bulk vertex operators:
\begin{equation}\label{vertex}
V_{m,n}=\int_{\Sigma} \d x \sqrt{g} e^{(1-m)2\pi \Phi/\hbar} e^{\i n \Phi}\,.
\end{equation}
We observe that our trumpet-creating vertex operators \eqref{2.23} are precisely such $(2,b)$ degenerates. In terms of the Liouville primary, these operators (which create physical states in our theory) have a label
\begin{equation}
    \alpha_{\text{state}\,b}=\frac{Q}{2}-b \frac{b_\text{CFT}}{2}
\end{equation}
We may also consider the set of physical operators acting on our Hilbert space of section \ref{sect:trumpets}, which ought to be invariant under $\Phi\to \Phi+2\pi$, at least in the minisuperspace approximation. They are the $(1,1+b)$ degenerates, which indeed are pure phases. So, the physical operators in sine dilaton gravity are
\begin{equation}
    V_{\text{operator}\,b}=\int_\Sigma\d x\sqrt{g}\,e^{\i(b+1)\Phi}\,,\quad \alpha_{\text{operator}\,b}=-b \frac{b_\text{CFT}}{2}\,\label{6.11}.
\end{equation}
The reason to label these with $b$ too will become clear below. Notice that, in sine dilaton gravity, if we consider a $\chi=0$ surface such as a cylinder, with as many of these operators on it as we want, that this configuration has an exact zero mode symmetry $\Phi(x)\to \Phi(x)+2\pi$.\footnote{In gravity the distinction between states and operators arises due to the fact that operators that create physical states need to prepare invariant wavefunctions when integrated on a disk, in which case the disk action provides an extra $e^{2\pi \Phi}$ whereas operators need to prepare gauge invariant functions when integrated on a cylinder, in which case the gravitational action does not contribute $\Phi$ dependence because $\chi=0$.}

The following is not to be interpreted as precise claims, but rather as speculative comments.

The fact that all the interesting vertex operators in our theory are degenerates suggest a relation between sine dilaton gravity and generalized minimal models \cite{Zamolodchikov:2005fy,Dotsenko:1984ad,Belavin:2005yj}. (Non-generalized) minimal models appear when $b_\text{CFT}^2=p/q$ with $p$ and $q$ being coprime integers, then the operator spectrum consists of primaries $\mathcal{O}_{n,m}$, characterized by the ranges $1 \leq m <q$ and $1 \leq n <p$. For general $b^2_\text{CFT}$ there still exists a similar structure known as generalized minimal models. The primary operator content of these is given by the full set of degenerate fields $\mathcal{O}_{m,n}$ in the Kac table \eqref{alpha_deg} with $(m,n)$ unbound integers, which is akin to what we found in sine dilaton gravity. In this context, a reduction of the full spectrum of timelike Liouville theory to a discrete set of primaries was achieved in \cite{Kapec:2020xaj} by compactifying the Liouville field. This suggests a mechanism closely resembling a periodic identification of the dilaton. Still, in sine dilaton gravity we observe only one tower $b$ of physical operators, whereas in the generalized minimal string there are two towers $(m,n)$. We believe we should restrict the generalized minimal model further to the spectrum \eqref{6.11} of operators in sine dilaton gravity by imposing periodicity of the dilaton. This makes sense, as the $(1,b+1)$ degenerates form a closed algebra under OPE \cite{Dotsenko:1984ad}. This ad hoc restriction breaks the usual $b_\text{CFT}\to 1/b_\text{CFT}$ symmetry, which indeed does not exist in our quantization of sine dilaton gravity \cite{Blommaert:2024whf}.

We think that there is a natural pairing between the vertex operators $V_{\text{operator}\, b}\leftrightarrow V_{\text{state}\,b}$, such that (in practice) multi-boundary amplitudes are computed using correlators of operators $V_{\text{operator}\, b}$. Before motivating this, let us give a piece of evidence that this could be on the right track. Namely, we believe that in generalized minimal string theory the sphere three point function equals the answer from the finite cut matrix integral \eqref{6.3volum}:
\begin{equation}
    \average{V_{\text{operator}\, b_1}V_{\text{operator}\, b_2}V_{\text{operator}\, b_3}}=\frac{1+(-1)^{b_1+b_2+b_3}}{2}\,.\label{6.12}
\end{equation}
To compute this, one multiplies the relevant DOZZ structure constants \cite{Zamolodchikov:2005fy,Dotsenko:1984ad}. Equation (2.38) in \cite{Collier:2024kwt} is a useful representation. This expression has poles at
\begin{equation}
    \alpha_1+\alpha_2+\alpha_3=Q\chi/2-n b_\text{CFT}-m/ b_\text{CFT}\,,\quad \chi=2\,.\label{6.13}
\end{equation}
with $\alpha$-independent residues, which we believe are to be interpreted as the relevant amplitude for sine dilaton gravity. Indeed, solutions to the WdW equation in sine dilaton gravity are also related to those in ordinary Liouville gravity by taking the residues of their poles, see section \eqref{subsect:ungworm} and \cite{Blommaert:2024whf}. We suspect that this is a general feature which persists for higher topologies if we specialize $\alpha$ to the degenerate values \eqref{6.11}. The pole structure in \eqref{6.13} gives the selection rule
\begin{equation}
    b_1+b_2+b_3=2 n\,,
\end{equation}
which we indeed see in \eqref{6.12} and \eqref{6.3volum}.

But then why does the gravity calculation of the trumpet, as implemented by the vertex operator given in \eqref{2.23}, seem like it involves the $(2,b)$ degenerate, which are not part of this spectrum \eqref{6.11}? We believe we can gain intuition by thinking about the Coulomb gas formalism \cite{Dotsenko:1984nm, Dotsenko:1985hi, Dotsenko:1984ad}. This approach involves expanding the Liouville interaction term and treating it perturbatively. Integration over the Liouville zero mode produces poles in the correlation functions, which occur precisely when the operators take degenerate values
\begin{equation}\label{pole}
\sum_{i}\alpha_i=Q \chi/2-n b_\text{CFT}\,.
\end{equation}
This misses the tower of $m$ poles discussed in \eqref{6.13}. Simply using operators \eqref{6.11} is is impossible to satisfy this selection rule unless $\chi=0$. We believe this is analogous to the statement that global shifts $\Phi(x)\to \Phi(x)+2\pi$ in sine dilaton gravity are only an obvious Lagrangian symmetry if $\chi=0$. What we are facing is the usual problem with taking the Liouville Lagrangian too serious. Indeed, even though this is less obvious, the complete formula of the Coulomb gas involves \emph{two} types of screening charges: $\exp(2b_\text{CFT} \phi)$ \emph{and} $\exp(2 \phi/b_\text{CFT})$. The second screening charge is responsible for $m\neq 0$ and for allowing topology change. In gravity it corresponds with including powers of $e^{2\pi m \Phi/\hbar}$. In general, these create $(m+1)$-cover singularities or crotches for $m=1$ \cite{Blommaert:2023vbz,Louko:1995jw}. It is not surprising to find that such operators play a role in allowing topology change. 

The combination of a trumpet \eqref{2.23} and a crotch behaves indeed as a physical puncture operator \eqref{6.11}. Screening charges allow us to make sense of a sphere with $n$-geodesic boundaries in sine dilaton gravity, as $n-2$ trumpets may be screened to become punctures, which do make sense on the cylinder. More generally, crotches can eat up the Euler character on higher genus such as to be consistent with a $\Phi(x)\to \Phi(x)+2\pi$ symmetry. It is remarkable that the path integral provides these screening charges. As a sanity check, shifting $\Phi(x)\to \Phi(x)+2\pi$ in the sine dilaton path integral with operators \eqref{6.11} and $m$ screening charges leads to a constraint $\chi+m=0$ on the number of crotches $m$; the same relation that follows from \eqref{6.13}. In summary, we believe that the naive obstruction of sine dilaton gravity with $\Phi(x)\to \Phi(x)+2\pi$ and topology change is similar to why we used to say topology change in Lorentzian JT gravity is impossible \cite{Usatyuk:2022afj,Blommaert:2023vbz}. Both are in fact possible.\footnote{The mechanism by which crotches are ``provided'' by the gravitational path integral was explained in \cite{Blommaert:2023vbz}.} We intend to flesh out this interpretation in more detail in the future. 

If such an interpretation is correct, one could consider deformations that preserve the symmetry\footnote{See also \cite{Giribet:2024men}.}
\begin{equation}
    I\to I+\int\d x\sqrt{g}\,\sum_{n=1}^\infty f(n)\, e^{\i n \Phi}\,.
\end{equation}
These would correspond with a gas of trumpets and in the matrix integral with exponentials of $O_\text{G}(b)$ \eqref{7.1OG}. Can one match classical thermodynamics on both sides?

\subsection{Towards observational no-boundary state}\label{subsect:nonoboundary}
To end this work, we briefly return to the decomposition of the sine dilaton gravity disk into trumpets \eqref{eq:okudecomp}
\begin{equation}
    Z_\text{disk}(\beta)=\sum_{b=-\infty}^{+\infty}\psi_\text{disk}(b)\, Z_\text{trumpet}(b,\beta)\,.
\end{equation}
The wavefunction $\psi_\text{disk}(b)$ has the geometric interpretation of the gravitational path integral ending on an AdS$_2$ geodesic (the half-sphere):
\begin{equation}
    \psi_\text{disk}(b)=\,\,\begin{tikzpicture}[baseline={([yshift=-.5ex]current bounding box.center)}, scale=0.7]
    \pgftext{\includegraphics[scale=1]{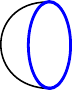}} at (0,0);
    \draw (0.9,0) node {\color{blue}$b$};
  \end{tikzpicture}
\end{equation}
The normalizable wavefunction which we obtained is a simple Gaussian \eqref{2.32}
\begin{equation}
    \psi_\text{disk}(b)=\cos(\pi b/2)q^{b^2/4+b/2}=\i q^{(b+1+\i \pi/\abs{\log q})^2/4}-\i q^{(b+1-\i \pi/\abs{\log q})^2/4}\,.\label{6.19psi}
\end{equation}
This replaces the usual expression in double-scaled matrix models where the Gaussians would collapse to delta functions. Unlike the deltas, this is a normalizable wavefunction, which is preferable in cosmology.\footnote{This also results in a finite sphere amplitude for sine dilaton gravity. This will be discussed elsewhere.} But what do these Gaussians mean? An educated guess is that this represents the trace over an auxiliary Hilbert space of an observer coupled to gravity \cite{Chandrasekaran:2022cip}. We imagine states $\ket{b,m}_\text{obs}\otimes \ket{b}_\text{grav}$ that are entangled in a density matrix which is diagonal in $b$, and with a Hamiltonian for both systems. There are then two contexts (at least) where it might be natural to identify $\psi_\text{disk}(b)$ with a trace in the observer Hilbert space (for fixed quantum number $b$).
\begin{enumerate}
    \item In 2d quantum gravity one interpretation suggests an observer behind the black hole horizon, like an EOW brane with actually quantized dynamical degrees of freedom. It seems natural to think about $b$ as a time variable for this observer, as the usual equation $\hbar b=2\pi\i$ fixes the boost angle around the horizon to $2\pi$. This boost angle is (in a sense) a diff invariant intrinsic Euclidean time variable for the horizon. It seems then that for sine dilaton gravity the state of the observer is not one of fixed time. This is indeed known to make black hole entropy deviate from the QES formula \cite{DeVuyst:2024uvd}. But then what makes this particular clock state ``special''?
    \item In 3d quantum gravity, sine dilaton gravity has a (rather well understood) interpretation as pure dS$_3$ quantum gravity or $SL(2,\mathbb{C})$ CS theory \cite{Gaiotto:2024kze,Verlinde:2024znh,Collier:2025lux}. $Z_\text{disk}(\beta)$ computes a partition function of the static patch. The holographic boundary is a tube surrounding the cosmological horizon. We do not know whether the tube is outside or inside. We do not know the boundary conditions on this tube, but we do know that they can be prepared by some linear combination of Wilson lines on the horizon (which create conical defects around them), as this forms a basis. Interestingly, in this interpretation the CS description of $Z_\text{trumpet}(b,\beta)$ involves an additional Wilson line in a representation labeled by $b$, which follows the worldline of the static patch observer\footnote{We thank H. Verlinde for past discussions on this point. This 3d dS geometry is built up from the sine dilaton topology by propagating the 2d geometry along the timelike isometry of the 3d static patch \cite{Gaiotto:2024kze,Verlinde:2024znh,Collier:2025lux}. This is a type of 2d-3d Cauchy slice holography where in addition the 2d theory has a 1d boundary dual near the horizon, in our case DSSYK. We drew the Penrose diagram of the Lorentzian static patch of 3d dS.}
    \begin{equation}
        Z_\text{disk}(\beta)=\sum_{b=-\infty}^{+\infty}\psi_\text{disk}(b)\,\,\begin{tikzpicture}[baseline={([yshift=-.5ex]current bounding box.center)}, scale=0.7]
    \pgftext{\includegraphics[scale=1]{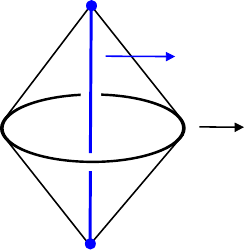}} at (0,0);
    \draw (5,0.3) node {hologram near};
    \draw (5,-0.4) node {cosmological horizon};
    \draw (-0.1,0) node {\color{blue}$b$};
    \draw (3.5,1.2) node {\color{blue}observer worldline};
    \draw (-2.2,-0.8) node {$\beta$};
  \end{tikzpicture}
    \end{equation}
    This is a massive particle in dS$_3$ with mass determined by the irrep $b$. The usual equation $\hbar b=2\pi \i$ is saying that the opening angle around the observer is $2\pi$. Thus this is the shrinkable boundary (with no mass). Apparently in our theory the mass is Gaussian distributed. But what makes this special? A semiclassical understanding of this duality to CS theory seems desirable, for instance what are the boundary conditions on the holographic boundary near the horizon, and where is it located?
\end{enumerate}
We stress that, at least from the fully-open channel quantization, we understood what makes the disk ``special'': the disk amplitude involves using the appropriate tracial state in the open channel \cite{Blommaert:2024whf}. Here, in the closed channel, however, it remains an open question what selects $\psi_\text{disk}(b)$. 

We give two rewritings of $\psi_\text{disk}(b)$ which have intuitive quantum mechanical interpretation as traces in a putative observer Hilbert space, hoping that these rewritings might help understand what is going on.
\begin{enumerate}
    \item Consider an SL$(2,\mathbb{R})$ finite dimensional representation $j=b/2$ with $C(j)=j(j+1)$. Using the first equation in \eqref{6.19psi}, one recognizes a twisted trace with a Euclidean time evolution $\beta_\text{obs}=\hbar/2$:
    \begin{equation}
        \psi_\text{disk}(b)=\Tr_j q^\mathbf{C}(-1)^{\mathbf{J}_z}\,.
    \end{equation}
    \item Consider the degenerate Virasoro matter primary \eqref{6.11} which we believe to be naturally paired with a geodesic boundary $b$. Consider now the Virasoro character associated with general matter degenerates \eqref{alpha_deg} (unfortunately the symbols $q$ and $b$ were already used in this work)
    \begin{equation}
        \text{ch}_{n,m}(\tau)=\frac{1}{\eta(\tau)}q_\text{CFT}^{-\frac14 \left(nb_\text{CFT}+m/b_\text{CFT}\right)^2}-\frac{1}{\eta(\tau)}q_\text{CFT}^{-\frac14 \left(nb_\text{CFT}-m/b_\text{CFT}\right)^2}\,,\quad q_\text{CFT}=e^{2\pi\i\tau}
    \end{equation}
    We then recognize the wavefunction as this character in a degeneration limit of the torus surface\footnote{This degeneration limit seems to be related with an independent observation by H. Verlinde about shrinkable boundary conditions in CS theory in an unusual channel \cite{hermanstrings}.}
    \begin{equation}
        \psi_\text{disk}(b)=\text{ch}_{b+1,1}(\tau=-1/2)\,.
    \end{equation}
    The identification $q_\text{CFT}=e^{-\i \pi}$ may vaguely suggest some relation with the usual dS temperature.
\end{enumerate}

\appendix

\section*{Acknowledgments}
We thank Xuchen Cao, Yasunori Nomura, Chang-Han Chen, Daniel Jafferis, Victor Ivo and Herman Verlinde for discussions. AB was supported by a Marvin L. Goldberger Membership, the US DOE DE-SC0009988 and by the Ambrose Monell Foundation. AL is supported by the Heising-Simons foundation under grant no. 2023-4430 and the Packard Foundation. KP was supported by the US DOE DE-SC0011941. TM and JP acknowledge financial support from the European Research Council (grant BHHQG-101040024). Funded by the European Union. Views and opinions expressed are however those of the author(s) only and do not necessarily reflect those of the European Union or the European Research Council. Neither the European Union nor the granting authority can be held responsible for them.

\section{Fixed acceleration branes in JT gravity}\label{app:details}
We discuss how to canonically quantize constant acceleration branes in JT gravity. We then apply this result to momentum branes in Sine-dilaton gravity. 

In JT gravity, constant extrinsic curvature branes obey the boundary conditions 
\begin{align}
    K = K_0\,\quad \partial_n \Phi - \Phi K_0 = m\,,
\end{align}
where $m$ is the parameter conjugate to the length of the curve. To discuss constant acceleration branes, it will be helpful to work with embedding coordinates for AdS$_2$. Using these coordinates, AdS can be described as the hyperboloid in three-dimensional Minkowski
\begin{align}
    -Y_{-1}^2 - Y_{0}^{2} + Y_1^2 = -1\,,\quad \d s^2 = -\d Y_{-1}^2 - \d Y_{0}^2 + \d Y_1^2.\,.
\end{align}
We can describe the embedding space coordinates in terms of both AdS-Rindler and global coordinates as
\begin{align}
    Y_{-1} = \cosh(\rho) = \frac{\cos(T)}{\sin(\sigma)}\,\quad Y_0 = \sinh(\rho) \sinh(t) = \frac{\sin(T)}{\sin(\sigma)}\,\quad Y_1 = \sinh(\rho) \cosh(t) = -\cot(\sigma)\,.
\end{align}
The induced metric on the hyperboloid using these coordinates is (for AdS-Rindler respectively global AdS)
\begin{align}
    \d s^2 = -\sinh^2(\rho) \d t^2 + \d\rho^2= \frac{-\d T^2 + \d\sigma^2}{\sin^2(\sigma)}\,.
\end{align}

In JT gravity, the dilaton solutions can be written as
\begin{align}
V \cdot Y = \Phi\,,\quad V^2 = -\Phi_h^2\,.
\end{align}
Now, following Appendix A of \cite{Maldacena:2016upp}, we can describe AdS curves of constant acceleration by the equation
\begin{align}\label{eqn:constantKcurve}
    Q \cdot Y = K_0\,\quad Q^2 = 1 - K_0^2\,,
\end{align}
where $K_0$ will turn out to be the extrinsic curvature of the slice. Note that the boosts of the ambient Minkowski form the group $SO(2,1)$, which acts as the group of isometries on AdS$_2$. These are treated as gauge symmetries, so we can fix a convenient gauge where $V^a = (\Phi_h, 0 ,0)$. Then the dilaton solution takes the form
\begin{align}
    \Phi = \Phi_h \frac{\cos(T)}{\sin(\sigma)}.
\end{align}
There is an $SO(1,1)$ subgroup that is left un-fixed by the choice for $V$. We can use this to further fix the form of $Q$. Parameterizing the unfixed $Q$ as
\begin{align}
    Q^a = \sqrt{1-K_0^2} (\alpha \cos(\theta), \alpha \sin(\theta), \sqrt{1+\alpha^2})\,,
\end{align}
we see that by doing a boost on the final two components we can set the second component to zero. This amounts to setting $\theta = 0$. Here $\alpha$ is a constant that will be fixed in terms of $K_0$ and $m$ in \eqref{eqn:alphavsmk}. 

From \eqref{eqn:constantKcurve} we see that the Rindler coordinates for the constant acceleration surface obey
\begin{align}
    -\alpha \cosh(\rho) + \sqrt{1+\alpha^2} \sinh(\rho) \cosh(t) = \frac{K_0}{\sqrt{1-K_0^2}}=\kappa\,.
\end{align}
This can be solved for $\rho(t)$\footnote{For $t \to \pm \infty$, $\rho \to 0$ as expected for constant acceleration curves, which cross the horizon. If we continue into Euclidean time, then indeed these curves reach the boundary when $\cos(\tau) = \alpha/\sqrt{1+\alpha^2}$. Note that for $|K_0|<1$, $|\kappa| > 1$.}
\begin{align}\label{eqn:rhovstime}
    \rho(t) = \log \frac{\sqrt{\kappa^2 -\alpha^2 + (1+\alpha^2) \cosh^2(t)}+ \kappa}{\sqrt{1+\alpha^2} \cosh(t) - \alpha} \,.
\end{align}
Similarly in global coordinates one finds
\begin{align}
    \sin(\sigma(T)) = - \frac{\alpha \kappa \cos(T) \pm \sqrt{(1+\alpha^2) (1+\alpha^2 + \kappa^2 - \alpha^2 \cos^2(T))}}{1+\alpha^2 + \kappa^2}\,.
\end{align}
The normal to this curve $\rho-\rho(t)=0$ is
\begin{align}\label{eqn:unitnormal}
    n^{\mu} = \frac{1}{\sqrt{\sinh^2(\rho) - \rho'(t)^2}} \left( \frac{\rho'}{\sinh(\rho)}, \sinh(\rho)\right).
\end{align}
Indeed using $K = \nabla_{\mu} n^{\mu}$ and \eqref{eqn:rhovstime} one checks that this curve has constant curvature $K = K_0$. Using \eqref{eqn:unitnormal} and $n^{\mu} \partial_{\mu} \Phi - \Phi K_0 = m$, we can eliminate $\alpha$ to completely determine the trajectory of the brane
\begin{align}\label{eqn:alphavsmk}
    \alpha = \frac{m}{\Phi_h\sqrt{1-K_0^2}}.
\end{align}

For the purpose of canonical quantization, as in \cite{Gao:2021uro} we need the lengths of geodesics ending normally on this brane. The charge $Q^a = \sqrt{1-K_0^2}\ (\alpha, 0 , \sqrt{1+\alpha^2})$ is found by performing a boost from $\alpha = 0$. One can then solve for the geodesic length in that frame and then use the isometries to move back to $\alpha \neq 0$, noting that the isometries do not change the geodesic length. When $\alpha = 0$, the curve takes the simple form in global coordinates
\begin{align}
    \sin(\sigma(T)) = \pm \frac{1}{\sqrt{1+\kappa^2}} = \pm \sqrt{1-K_0^2}.
\end{align}
Global geodesics are constant $T$ curves so the geodesic length can be compute by integrating
\begin{align}\label{eqn:geolengthnontransformed}
    L=\int_{\sigma(T)}^{\sigma_0} \frac{\d\sigma}{\sin(\sigma)} = \frac{1}{2} \bigg(\log \frac{1-\cos(\sigma_0)}{1+\cos(\sigma_0)} - \log \frac{1-K_0}{1+K_0} \bigg).
\end{align}
Here $\sigma_0$ is the location where the geodesic hits the holographic boundary particle, and determining it is our remaining to-do. We can Lorentz transform in the ambient Minkowski to the $\alpha\neq 0$ frame using
\begin{align}
    Q_0^a \Lambda^b_{a} = Q^a = \sqrt{1-K_0^2}\ (\alpha , 0 ,\sqrt{1+\alpha^2})\,,\quad \Lambda_a^b = 
    \begin{pmatrix}
\sqrt{1+\alpha^2} & 0 & \alpha \\
0 & 1 & 0 \\
\alpha & 0 & \sqrt{1+\alpha^2}
    \end{pmatrix}\,.
\end{align}
This is the same as the transformation in \cite{Gao:2021uro} and so we can use their equation (2.26):
\begin{align}
    \cos \sigma_0 \to - \frac{\sqrt{1+\alpha^2} \cos(\sigma_0) - \alpha \cos(T_0)}{\sqrt{\sin^2(T_0) + (\sqrt{1+\alpha^2} \cos(T_0) -\alpha \cos(\sigma_0))^2}}
\end{align}
After subtracting the divergent piece, one finds the renormalized length
\begin{align}
    L = \log \frac{2 (\alpha + \sqrt{1+\alpha^2}\sec(T_0))}{\Phi_h} - \frac{1}{2}\log \frac{1-K_0}{1+K_0}\,.
\end{align}
Remembering equation \eqref{eqn:alphavsmk} and using equation 2.16 in \cite{Gao:2021uro} that boundary time $w$ satisfies $\tan(T/2)=\tanh(w)$ we finally obtain equation \eqref{eqn:Ltsolution} in the main text:
\begin{align}
    L = \log \frac{m + \sqrt{(1-K_0^2)\Phi_h^2 + m^2} \cosh(w)}{\sqrt{1-K_0^2}\Phi_h^2} - \frac{1}{2}\log \frac{1}{4}\frac{1-K_0}{1+K_0}\,.
\end{align}
For the quantization of branes in sine dilaton this is what we needed to derive.

Since gravity is still dynamical on the brane, the contribution to the AdM Hamiltonian of the system from the brane is zero. The relationship between the AdM Hamiltonian and $\Phi_h$ is the same as in \cite{Gao:2021uro}, namely $H=\Phi_h^2$. Shifting the length by a constant leads to the generic JT gravity brane Hamiltonian
\begin{align}
    H_{m K_0} = \frac{1}{1-K_0^2}\bigg( \frac{P^2}{4} + me^{-L}+e^{-2L}\bigg)\,.
\end{align}
This is identical to the geodesic system of \cite{Gao:2021uro} up to energy rescaling. This reflects the fact that, in JT gravity, the conjugate to length is $m$ in the closed channel, so curvature is a gauge choice.

\section{Details on deriving trumpets from brane quantum mechanics} \label{sec:trumpetdetails}
In this appendix we compute the thermal partition function using the semi-open brane Hamiltonians \eqref{3.38hh}. We aim to extract the trumpet wavefunction by Fourier transforming from $\Phi_h$ to $b$. To proceed, we expand into fixed length eigenstates to write the thermal trace as
\begin{align}\label{eqn:thermaltrace}
    \Tr e^{-\beta \mathbf{H}_{\mu \bmu}/\hbar} = \sum_{n=0}^{\infty} \int_0^\pi \d\theta\, \big(e^{\pm 2\i\theta};q^2\big)_\infty \braket{\mu, \bmu|\theta} e^{\beta \cos(\theta)/\hbar}\, \Psi_n(\theta;\mu,\bmu) \Psi_n(\theta;\mu, \bmu),
\end{align}
where the wavefunctions are Al-Salam-Chihara polynomials that solve \eqref{eqn:ASCpoly}
\begin{align}\label{eqn:psiQ}
    \Psi_n(\theta;\mu, \bmu) = \frac{Q_{n}(\cos(\theta_1)|q^2)}{\big(q^2,AB;q^2)_{n}^{1/2}}\,,
\end{align} 
ans the constants $A,\ B$ are given in \eqref{eqn:ABtheta}. The normalization \eqref{eqn:ASCnormalization} is
\begin{align}
    \int_0^{\pi} \d\theta\, \big(e^{\pm 2\i\theta};q^2\big)_\infty \braket{\mu,\bmu|\theta} \Psi_n(\theta;\mu, \bmu) \Psi_m(\theta;\mu, \bmu) =\delta_{nm},
\end{align}
The overlap $\braket{\mu, \bmu| \theta}$ is given in \eqref{eqn:mubmuthetaoverlap}. Note that the Al-Salam polynomials in this normalization obey a completeness relation of the form
\begin{align}
    \sum_{n=0}^{\infty} \Psi_n(\theta; \mu, \bmu) \Psi_n(\theta'; \mu, \bmu) = \frac{\delta(\theta - \theta')}{\big(e^{\pm 2\i\theta};q^2\big)_\infty \braket{\mu, \bmu| \theta}}, 
\end{align}
and so equation \eqref{eqn:thermaltrace} is formally divergent. Nevertheless, by Fourier transforming in $\Phi_h(\mu, \bmu)$, we can extract a non-divergent trumpet wavefunction, as was done in \cite{Okuyama:2023byh, Gao:2021uro}. To proceed, it will be useful to write the wavefunctions in the momentum (conjugate to $n$) basis as
\begin{align}
    \frac{Q_n(\cos \theta|q^2)}{(q^2;q^2)_n} = \int_0^{2\pi} \d p_1 e^{-\i p_1n} \frac{(A e^{\i p_1}, B e^{\i p_1};q^2)_{\infty}}{(e^{\pm \i \theta + \i p_1};q^2)_{\infty}}.
\end{align}
Transforming this integral to $z_1 = e^{ip_1}$ variables, it is actually fairly easy to see that these wavefunctions vanish for $n \in \mathbb{Z}_{<0}$. This is because, when $n<0$, there is no pole at the origin in $z_1$ space. Thus, when we compute the sum over $n$ in the thermal trace we are free to extend the sum over $n$ to all integers. 

Now computing the (formally divergent) trace using the wavefunctions \eqref{eqn:psiQ} one obtains 
\begin{align}
    &\Tr e^{-\beta \mathbf{H}_{\mu \bmu}/\hbar} =\sum_{n=-\infty}^\infty (q^2;q^2)_n(e^{ip_- + ip_+}q^{2n};q^2)_{\infty}\int_0^\pi\d\theta\, e^{\beta \cos \theta}\frac{\big(e^{\pm 2\i\theta};q^2\big)_\infty }{\big(e^{\pm\i\theta+\i p_{\pm}};q^2\big)_\infty}\\\nonumber&\qquad \int_{0}^{2\pi}\d p_1\,e^{-\i p_1 n}\frac{\big(e^{\i p_1+\i p_{\pm}};q^2\big)_\infty}{\big(e^{\i p_1\pm \i \theta};q^2\big)_\infty}\int_{0}^{2\pi}\d p_2\,e^{-\i p_2 n}\frac{\big(e^{\i p_2+\i p_{\pm}};q^2\big)_\infty}{\big(e^{\i p_2\pm \i \theta};q^2\big)_\infty}\,.
\end{align}
In order to slightly lighten the notation we introduced
\begin{equation}
    p_{\pm} = \overline{\alpha} \pm \alpha\,.
\end{equation}
Introducing $z=e^{\i p_+}=e^{-\i \Phi_h}$ and $z_{1,2}=e^{\i p_{1,2}}$, we can then write the Fourier transform in $p_+$ as \footnote{We have also expanded this expression, using the relations
\begin{equation}
    \big(q^2;q^2\big)_n=\sum_{m_1=0}^\infty \frac{\big(q^2;q^2)_\infty}{(q^2;q^2)_{m_1} }q^{2m_1}q^{2n m_1}\ , \quad \quad
    (e^{ip_-+ip_+}q^{2n};q^2)_\infty = \sum_{m_2=0}^\infty \frac{q^{m_2(m_2-1)} e^{\i(p_1+p_+)m_2 + 2nm_2}}{(q^2;q^2)_{m_2}}.\, .
\end{equation}}
\begin{align}\label{eqn:fulltrumpetexpanded}
   & Z_\text{trumpet}(b,\beta)=\frac{1}{2\pi\i}\oint\frac{\d z}{z}\,z^{-b}\nonumber\\
    &\sum_{n=-\infty}^\infty \sum_{m_1, m_2=0}^\infty \frac{z^{m_2}\big(q^2;q^2)_\infty q^{2m_1+ 2n (m_1+m_2) + i p_- m_2 + m_2(m_2-1)}}{(q^2;q^2)_{m_1} (q^2;q^2)_{m_2}} \int_0^\pi\d\theta\, \frac{\big(e^{\pm 2\i\theta};q^2\big)_\infty}{\big(ze^{\pm\i\theta};q^2\big)_\infty\big(e^{\pm\i\theta+\i p_{-}};q^2\big)_\infty}\nonumber \\
    &\qquad\frac{1}{2\pi\i}\oint \frac{\d z_1}{z_1}\,z_1^{-n}\frac{\big(z z_1;q^2\big)_\infty (z_1e^{ip_-};q^2)_\infty}{\big(z_1 e^{\pm \i \theta};q^2\big)_\infty}\frac{1}{2\pi\i}\oint \frac{\d z_2}{z_2}\,z_2^{-n}\frac{\big(z z_2;q^2\big)_\infty (z_2e^{\i p_-};q^2)_\infty}{\big(z_2e^{\pm \i \theta};q^2\big)_\infty}\,.
\end{align}
The contours for $z_1$ and $z_2$ lie slightly inside the unit circle, and close to the outside. The poles are at
\begin{equation}\label{eqn:z1z2poles}
    z_1=e^{\pm \i \theta}q^{-2r_1}\,,\quad z_2=e^{\pm \i \theta}q^{-2r_2}\,.
\end{equation}

Now, there are three orders of integration one could follow in \eqref{eqn:fulltrumpetexpanded}. Suppose we first do the $z_1$ and $z_2$ contour integrals for generic $z$, picking up the poles in \eqref{eqn:z1z2poles}. This gives generically finite residues. One can then sum over $n$ using
\begin{equation}
    \sum_{n=-\infty}^{+\infty}(q^{2(m_1+m_2)}/z_1z_2)^n=\sum_{n=-\infty}^{+\infty}q^{2n(m_1+m_1+r_1+r_2)}e^{\pm \i \theta n\pm \i \theta n}
\end{equation}
This vanishes unless $m_1=m_2=r_1=r_2=0$ and the signs of the phases are opposite - in which case we find $\delta(0)$. This is the analytic continuation of the Dirac comb formula, which is allowed to be used in this distributional sense.\footnote{For an earlier application of such an equation, see e.g. appendix C of \cite{Mertens:2014nca}, where it is noted that one has to be careful with such formulas as e.g. the general Poisson resummation formula is more subtle to interpret in the complex plane.} This is the correct answer for the annulus with a circular EOW brane, which is divergent because of large $n$ asymptotics of the wavefunctions (or because continuous quantum mechanical systems have divergent partition functions), see also \cite{Gao:2021uro}.

The other orders of integration are more interesting. We first do the $z$ integral, picking up poles at
\begin{equation}\label{eqn:poles}
    z=e^{\i\theta}q^{-2r}\ \text{and } z=e^{-i\theta}q^{-2r}\,.
\end{equation}
When $b>0$, we may close the contour outwards away from the origin to pick up the poles \eqref{eqn:poles}. We obtain\footnote{Note that the factor $\frac{(1;q^2)_\infty}{(q^{-2r};q^2)_\infty}$ is the ratio of two zeroes. What we mean by this notation is $\frac{(1;q^2)_\infty}{(q^{-2r};q^2)_\infty} \equiv \lim_{z \to 1} \frac{(z;q^2)_\infty}{(zq^{-2r};q^2)_\infty}$.}
\begin{align}
    &Z_\text{trumpet}(b,\beta)= \sum_{r=0}^\infty q^{2 r b} \sum_{n=-\infty}^\infty \int_0^\pi\d\theta\, \frac{e^{-\i b\theta}\big(e^{\pm 2\i\theta};q^2\big)_\infty}{\big(e^{2\i\theta}q^{-2r};q^2\big)_\infty\big(e^{\pm\i\theta+\i p_{-}};q^2\big)_\infty} \frac{(1,q^2)_\infty}{(q^{-2r};q^2)_\infty}\label{6.13bis}\\&\frac{1}{2\pi\i}\oint \frac{\d z_1}{z_1}\,z_1^{-n}\frac{\big(z_1e^{\i \theta}q^{-2 r};q^2\big)_\infty(z_1e^{\i p_-};q^2)_\infty}{\big(z_1 e^{\pm \i \theta};q^2\big)_\infty}\frac{1}{2\pi\i}\oint \frac{\d z_2}{z_2}\,z_2^{-n}\frac{\big(z_2e^{\i\theta}q^{-2 r};q^2\big)_\infty(z_2e^{\i p_-};q^2)_\infty}{\big(z_2e^{\pm \i \theta};q^2\big)_\infty}+(\theta \to -\theta)\, ,\nonumber
\end{align}
where we set $m_1 = m_2 = 0$ in \eqref{eqn:fulltrumpetexpanded}, since these are the only terms which survive the sum over $n$. Now to get a finite answer at fixed $b$, we bring the $n$ sum inside. The only effect at this point is to produce a delta function in $p_1 - p_2$. Picking up this delta function, we are left with
\begin{align}
    &Z_\text{trumpet}(b,\beta)= \sum_{r=0}^\infty q^{2 r b}
    \int_0^\pi\d\theta\, \frac{e^{-\i b\theta}\big(e^{\pm 2\i\theta};q^2\big)_\infty}{\big(e^{2\i\theta}q^{-2r};q^2\big)_\infty\big(e^{\pm\i\theta+\i p_{-}};q^2\big)_\infty} \frac{(1,q^2)_\infty}{(q^{-2r};q^2)_\infty}\nonumber\\&\qquad \frac{1}{2\pi\i}\oint \frac{\d z}{z} \frac{\big(ze^{\i \theta}q^{-2 r};q^2\big)_\infty(ze^{\i p_-};q^2)_\infty\big(e^{\i\theta}q^{-2 r}/z;q^2\big)_\infty(e^{\i p_-}/z;q^2)_\infty}{\big(z e^{\pm \i \theta};q^2\big)_\infty\big(e^{\pm \i \theta}/z;q^2\big)_\infty}+ (\theta \to -\theta)\,. \label{6.14}
\end{align}
If we deform the $z$-contour outwards away from the origin then, naively, we pick up single poles at the zeroes of the denominator in \eqref{6.14}, which occur at
\begin{align}
    z = e^{\pm i \theta}q^{-2m},\ m>0\,.
\end{align}
However, the numerator has zeroes precisely at these locations, so that these are not poles. There are, however, genuine poles in the integral at\footnote{Remember that the defining contour for the $z_1$ and $z_2$ integrals in \eqref{6.13bis} lies just inside the unit circle and so the poles in \eqref{eqn:specialpoles} are indeed picked up when we deform the $z$ integral outwards.}
\begin{align}\label{eqn:specialpoles}
    z = e^{i \theta}\ \text{and } e^{-i\theta}. 
\end{align}
At these locations, the denominator has a double zero but the numerator only has a single zero, leaving simple poles at $z = e^{\pm i \theta}$ in \eqref{6.14}. Picking up the residues at just these two poles gives a miraculously simple answer, which is completely independent of $p_-$
\begin{equation}
    Z_\text{trumpet}(b,\beta) = \frac{2}{\big(q^2;q^2\big)_\infty}\int_0^\pi\d\theta\,e^{-\i b\theta}\sum_{r=0}^\infty q^{2 r b}\,e^{\beta \cos(\theta)}+(\theta \to -\theta)\,.
\end{equation}
Doing the $\theta$ integral finally leads to\footnote{
We stripped off constants independent of $b, \theta$ or $\beta$ which can be absorbed in normalization of the $\ket{b}$ states in any case.}
\begin{equation}
    Z_\text{trumpet}(b,\beta)=\frac{1}{1-q^{2b}}\,I_b(\beta)\,.\label{7.14}
\end{equation}
Note that this same order of integration, where the $n$ sum is performed before the $z_1, z_2$ integrals, was used in \cite{Gao:2021uro}. The reason for brane redundancy from this perspective is that we ended up localizing on the residues of only the special set of poles in \eqref{eqn:specialpoles}. \eqref{7.14} reproduces our trumpet amplitude \eqref{2.28} from section \ref{subsect:finitecutoff} up to a Selberg determinant which can be absorbed in the normalization of the state $\ket{b}$. This completes the proof of equation \eqref{4.1zzzzzzz}, which was the goal of this appendix.

\bibliographystyle{ourbst}
\bibliography{Refs}

\end{document}